\documentclass[11pt]{article}

\usepackage{subfig}

\usepackage{amsfonts}
\usepackage{amssymb}
\usepackage{epsfig}
\usepackage{dcolumn}
\usepackage{color}
\usepackage{ mathrsfs }
\usepackage{graphicx}
\usepackage{caption}
\usepackage{subfig}

\makeatletter
\providecommand\phantomcaption{\caption@refstepcounter\@captype}
\makeatother

\def\R{\mathbb R}

\def\C{\mathbb C}

\advance \textheight by \topskip
\usepackage{amsmath,dsfont}
\usepackage[english]{babel}

\makeatletter
 \@addtoreset{equation}{section}
\makeatother

\usepackage[utf8]{inputenc}
\usepackage{amsmath,amssymb}
\makeatletter \@addtoreset{equation}{section}
\makeatother
\def\be{\begin{equation}}

\def\be{\begin{equation}}
\def\ee{\end{equation}}

\def\R{\mathbb R}

\usepackage{amsmath,amssymb}
\def\bea{\begin{eqnarray}}
\def\eea{\end{eqnarray}}
\def\barray{\begin{array}}
\def\earray{\end{array}}

\begin{document}

\title{
{\bf  $PT$-symmetric invisible defects and confluent
Darboux-Crum transformations }
}

\author{ 
{\small \textrm{\textup{\textsf{ 
Francisco Correa${}^{a,b}$, 
V\'{\i}t Jakubsk\'y${}^{c}$,
Mikhail S. Plyushchay${}^{d}$
}}} } \\
[15pt]
{\small \textit{ ${}^{a}$Leibniz Universit\"at Hannover, Appelstra\ss{}e 2, 30167 Hannover, Germany }}\\
{\small \textit{ ${}^{b}$Centro de Estudios Cient\'{\i}ficos (CECs), Arturo Prat 514, Valdivia, Chile}}\\
{\small \textit{${}^{c}$Department of Theoretical Physics, 
Nuclear Physics Institute, 25068  \v Re\v z, 
Czech Republic}}\\
{\small \textit{
${}^{d}$Departamento de F\'{\i}sica,
Universidad de Santiago de Chile, Casilla 307, Santiago 2,
Chile  }}\\
[10pt]
 \sl{\small{E-mails: 
correa@cecs.cl, jakub@ujf.cas.cz, mikhail.plyushchay@usach.cl}} 
}

\date{}

\maketitle

\begin{abstract}
We show that confluent Darboux-Crum transformations 
with emergent 
Jordan states 
are an effective tool for the design of optical 
systems governed by the Helmholtz equation under the paraxial approximation. 
The construction
 of generic, asymptotically real and periodic, $PT$-symmetric systems with local complex 
 periodicity defects is discussed in detail. We show how the decay rate of the defect is related 
 with the energy of the bound state trapped by the defect.  In particular, the bound states in 
 the continuum are confined by the periodicity defects with power law decay. We show that 
 these defects possess complete invisibility; the wave functions of the system coincide 
 asymptotically with the wave functions of the undistorted setting. The general results are 
 illustrated with explicit examples of reflectionless models and systems with one spectral gap. 
 We show that the spectral properties of the studied models are reflected by Lax-Novikov-type
  integrals of motion and associated  supersymmetric structures of bosonized and exotic nature.
\end{abstract}

\section{Introduction}

Propagation of light beams 
can 
mimic 
evolution of matter waves in quantum systems. 
This follows from the formal coincidence of  the Helmholtz equation for monochromatic light, 
propagating in magnetization free medium,  and the {stationary} Schr\"odinger equation in quantum mechanics \cite{Milonni}.
In this manner, behavior of quantum systems can be simulated 
by optical systems, and, vice versa, concepts common in optics can find their way to quantum settings.

Nowadays, there can be constructed optical materials where intensity of light is subject to a 
controlled gain and loss \cite{eo1}, \cite{eo2}, \cite{eo3}. 
Unusual optical properties of these systems are reflected by refractive index that acquires complex values. 
The perfect match between gain and loss prevents the light from being exponentially dimmed 
or brightened and it is reflected by invariance of the refractive index with respect to combined 
transformations of
space-inversion ($P$) and time-reversal ($T$). 
The light passing through such a material can exhibit some remarkable properties. 
Among them, let us mention  power oscillations or non-reciprocity of beam propagation that are 
caused by non-orthogonality of Bloch states \cite{po1}, \cite{po2}, \cite{po3}, \cite{nono1}, 
\cite{nono2}, violation of Friedel's law of Bragg scattering \cite{bp2}, unidirectional invisibility 
\cite{ui1}, \cite{ui2}, \cite{ui3}, 
\cite{ui4} or invisibility of defects in the periodic structure of 
optical crystals \cite{i1}, \cite{i2}.

The peculiar features of the $PT$-symmetric optical systems can be captured by  the Helmholtz 
(Schr\"odinger) equation with a non-hermitian Hamiltonian.
Such Hamiltonian operators 
have been studied extensively in quantum mechanics for last two decades, see \cite{BR}, \cite{MR1}, 
\cite{MR2} for review. 
It was understood that reality of their spectra occurs due to existence of an antilinear 
integral of motion that was identified with the $PT$ operator in most cases. 
The lack of hermiticity of the energy operator challenges probability interpretation for a
 quantum system; 
the scalar product has to be redefined 
\cite{benderZnojilmostafazadeh1}-\cite{benderZnojilmostafazadeh3} and its explicit form
 is usually exceedingly difficult to find. 

As the solutions of the Helmholtz equation correspond to a purely classical object 
(amplitude of the electric field), the hassle of the probability interpretation is avoided.
 In fact, the optical systems with balanced gain and loss open the door for realization of interesting phenomena 
predicted by non-Hermitian ($PT$-symmetric) quantum mechanics. 
One of them corresponds to the transition between exact and spontaneously broken $PT$ symmetry, where the behavior of 
a physical system at the vicinity of the exceptional spectral points was 
of the main 
interest \cite{bp1}, \cite{bp2}, \cite{bp3}.

The Helmholtz equation for linearly
polarized light acquires the form
\begin{equation}\label{Helmholtz}
	 \left(\partial_z^2+\partial_x^2+\frac{n^2(x)}{c^2}\omega^2\right)\Psi(x,z)=0,
\end{equation}
where $\omega$ is the frequency of monochromatic beam, 
and
$\Psi$ is electric field. The refractive index  $n(x)=n_0+\delta n(x)$, 
where 
$n_0>0$ and $|\delta n(x)|<<n_0$, 
is $PT$ symmetric, 
i.e. $\delta n(x)={\delta n(-x)}^{*}$. 
In dependence of its sign, the imaginary part of $\delta n(x)$ represents loss or gain.  
In the paraxial approximation, we set $\Psi(x,z)=e^{i\frac{n_0\omega}{c}z}\psi(x,z)$,
where $\psi(x,z)$ is an envelope function slowly varying in $z$. 
Neglecting the term with $\partial_z^2\psi$, the Helmholtz
equation reduces then to
\begin{equation}\label{sit1}
	 i\partial_z\psi(x,z)=[-\partial_x^2+V(x)]\psi(x,z),
\end{equation}
where $z$ denotes 
a
depth of propagation of light in the crystal while $x$ is 
a
transverse coordinate. The potential $V(x)$ is proportional to $\delta n(x)$, $V(x)\sim \delta n(x)$. 
{The form of physically acceptable solutions of (\ref{sit1}) depends on characteristics of the considered system; they can be square integrable in $x$, or quasi-periodic in $x$ when $V(x)$ is a periodic function. It is worth stressing that  
contrary to the
$PT$-symmetric quantum mechanics\footnote{ {The Hamiltonian $H$ with complex, $PT$ 
symmetric potential ceases to be self-adjoint with respect to the usual scalar product $(f,g)=\int_{-\infty}^{\infty} f^*(x)g(x)dx$. To recover physical relevance of the scalar product and the self-adjointness of the Hamiltonian, 
it is necessary 
to make a suitable redefinition of the former one, see e.g. \cite{MR1}, \cite{Bender:2002vv}.
 }
}, 
the quantity 
$\mathcal{P}(z)=\int_{-\infty}^{\infty}|\psi(x,z)|^2dx$ is physically relevant 
 as it denotes the 
 power of the beam $\psi(x,z)$
when measured in the depth $z$ within the crystal.
 Due to the complex refractive index, the power $\mathcal{P}(z)$ can be oscillating function in $z$ as the beam 
 propagates though the medium, see e.g. \cite{po2}.}

In the current article, we will have in mind primarily the framework of paraxial approximation 
described by (\ref{sit1}), {suitable for the low frequencies where the term $\partial_z^2\psi$ can be neglected}. 
Let us notice that another situation is also of interest in the literature, 
where no paraxial approximation is employed \cite{ui1}, \cite{ui3}. By separation of variables $\Psi(x,z)=e^{i\lambda z} \psi(x)$, the equation (\ref{Helmholtz}) 
reduces effectively to 
$	  \left(
	\frac{d^2}{dx^2}
	  +\frac{n^2(x)}{c^2}\omega^2-\lambda^2\right)\psi(x)=0.$
Here, the inhomogeneities of refractive index are developed in the direction of propagation of light, 
contrary to (\ref{sit1}) where the light propagates in normal direction to the inhomogeneities of $n(x)$.  
Our results 
will be applicable in this scheme as well. 

Supersymmetric quantum mechanics (SUSYQM) provides  powerful techniques for 
construction and analysis of quantum systems. It allows us to alter interaction terms of 
both exactly- and quasi-exactly solvable models without compromising their solvability properties.
It relates the scattering characteristics such as reflection and
transmission amplitudes  
of the two systems \cite{CKSsusy}. It can serve to find integrals of motion of the 
new systems; in the case of finite-gap and reflectionless systems, the nontrivial (Lax-Novikov) integrals can be identified in straightforward manner \cite{trisusy1,trisusy2, exotic1, exotic2}. 
The framework of SUSYQM can also be used very efficiently in the analysis of
 the soliton scattering, 
or to construct (multi-)soliton solutions of Korteweg-de Vries (KdV) and modified 
Korteweg-de Vries
equations
proceeding from the trivial solutions, see e.g. \cite{MPAA2013}, \cite{MPAA2014}.

SUSYQM relies on the intertwining operators.  These are frequently represented 
in terms of higher-order differential operators and identified as  Darboux-Crum
transformations known from the analysis of differential equations \cite{Matveev}. 
They are usually used to construct systems with altered number of discrete energies\footnote{They also can be used to eliminate some states  in the continuous part 
of the spectrum, see \cite{FMPT}.}. Requirement of regularity of a new system
imposes some restrictions
to the use and setup of Darboux-Crum
transformations.  Discrete
energies that lie below the spectral threshold of the original system can be added by 
the first order Darboux transformations. To generate discrete energy levels within a
 bounded gap  separating the spectral bands 
of an original, periodic system, Darboux-Crum
transformations of higher order 
{have} to be employed, see \cite{longwork}.

{Methods of quantum mechanics} provide efficient tools for analysis of optical systems.
 {They} have been used in the optics for a few decades \cite{optsusy1}-\cite{optsusy7}.  
In recent years, popularity of the supersymmetric approach witnesses growing 
interest in the context of the $PT$-symmetric optical devices \cite{susyPTopt1}-\cite{susyPTopt6}. 
It allows to construct solvable models
of optical crystals with reflectionless interfaces \cite{susyPTopt1},  the systems 
possessing unidirectional invisibility \cite{susyPTopt4}{, or periodic crystals} with 
invisible periodicity defects \cite{susyPTopt3}\footnote{{Invisibility of defects in 
the crystal structure means that the asymptotic form of the wave packet 
outgoing from the defect coincides with the wave packet that would 
propagate through the undistorted crystal. There, the reflection coefficient vanishes and the change in the phase factor reduces to the trivial one (modulo period of the potential), see e.g. \cite{susyPTopt3}.
} 
}.
 
Developing {further} the pioneering ideas of {von} Neumann and Wigner \cite{NeuWig},
a supersymmetric generalization  of the procedure 
for the construction of 
spherically symmetric scattering potentials 
that support bound states 
in the continuum was proposed in Refs. \cite{BICSukhatme}, \cite{Pursey}. 
General aspects of the technique, known in the literature as 
confluent Darboux(-Crum) transformation, have been analyzed in the series of 
papers \cite{CSUSY1,CSUSY2,CSUSY3,CSUSY4,CSUSY5}, see also refs. \cite{cc1,cc2}. 
The extension of the generalized framework to the systems defined on the 
whole real axis leads to the construction of non-Hermitian systems. Those 
were considered in the context of optics e.g. in \cite{BICoptics3}, \cite{BICoptics4}. 

Bound states in the continuum (BIC) can be observed experimentally in 
optical systems \cite{BICoptics1} and their theoretical 
investigation  in the context of $PT$-symmetric lattices was done,
for instance, in refs. \cite{BICoptics2}, \cite{BICoptics5}. The bound states
 located
on the threshold of the continuum spectrum of PT symmetric systems were 
considered in \cite{susyPTopt3}.

In the present work, by a systematic employment of SUSYQM, 
we will construct $PT$-symmetric optical systems 
where the refraction index is \emph{asymptotically real and periodic},
however, there are \emph{localized complex} periodicity defects.
We will show analytically that the decay rate of the defects is related to the 
energy of the bound states induced by them; the defects fall off as $x^{-1}$ 
or $x^{-2}$ for the systems with BIC, whereas exponential decay takes place 
for bound states associated with discrete energies.  

The work is organized as follows. In the
next section, we will present the main characteristics of the confluent 
Darboux-Crum transformations (also called double 
step Darboux transformations in the literature), the basic tool for construction of 
the systems with bound states in the continuum. We will explain its relation to
 the standard supersymmetric quantum mechanics and provide formulas for the 
 bound states and potential of the new system.
We will review basic properties of periodic systems in  section~\ref{BFsection},
 where we focus on the $PT$-symmetry of Bloch-Floquet states. In 
 section~\ref{PTGENsection}, we will focus on the $PT$-symmetric potentials
  that support bound states in the continuum. Results of the systematic analysis are
   illustrated on the examples in section~\ref{EXsection}. 
We discuss there, particularly, reflectionless systems that possess either
visible or invisible defects. 
 We consider one-gap systems that possess bound states in the continuum as well. In 
section~\ref{hifi}, we discuss some peculiar properties of reflectionless and
 finite-gap systems. In particular, we focus on the integrals of motion that 
 reflect spectral properties of the systems manifested in 
 two kinds
  of 
 associated superalgebras. One of them corresponds to a hidden bosonized 
 supersymmetry.
  It is based here on
  Lax-Novikov integrals mixed up with confluent Darboux-Crum
 transformations. 
The other one is an exotic hidden nonlinear supersymmetry 
 based 
  on
  the extended matrix Hamiltonian
 and containing the  
 extended number of 
integrals
  of 
   motion in comparison with a usual $N=2$ 
   supersymmetric structure. 
   The last section is devoted to 
discussion and outlook.

\section{Confluent Darboux-Crum transformation and Jordan states}

For any quantum system we obtain here a
partner  by employing a confluent
Darboux-Crum transformation, 
and observe the emergence of the
Jordan states  in {the} construction.
Such states will appear later in the structure of the
Lax-Novikov integrals controlling the invisibility  
of the $PT$-symmetric defects.

Consider a quantum system given by the Schr\"odinger Hamiltonian
\begin{equation}\label{A1}
	 H=-\frac{d^2}{dx^2}+V(x)\,.
\end{equation}
Let
$\psi_0(x)$ be a (physical or non-physical) 
eigenstate of eigenvalue $E_0$, 
\begin{equation}\label{A2}
	(H-E_0)\psi_0=0\,.
\end{equation}
We use it to introduce 
the first order differential operators
\begin{equation}\label{A3}
	A:=\psi_0\frac{d}{dx}\frac{1}{\psi_0}=\frac{d}{dx}-\frac{\psi_0'}{\psi_0}\,,\qquad
	A^{\sharp}:=-\frac{1}{\psi_0}\frac{d}{dx}\psi_0=-\frac{d}{dx}-\frac{\psi_0'}{\psi_0}\,.
\end{equation}
By definition, 
\begin{equation}\label{A4}
	A\psi_0=0,\  \qquad
	A^\sharp \frac{1}{\psi_0}=0\,.
\end{equation}
A linear independent from $\psi_0$ eigenstate of $H$
of the same eigenvalue $E_0$  we take in a form
\begin{equation}\label{A6}
	{\psi}_0{}^\flat(x)= \psi_0(x)
	\left(\int_{x_0}^x \frac{ds}{\psi^2_0(s)}+a\right)\,,\quad
	a\in \C\,,
\end{equation} 
where $x_0\in \R$ is some fixed point.
Notice that the Wronskian $W(f,g):=fg'-f'g$ of 
$\psi_0$ and ${\psi_0}^\flat$ is unit,
$W(\psi_0, {\psi_0}^\flat)=1$.
In what follows by ${\psi}^\flat(x)$ we shall denote a function 
associated with  a function $\psi(x)$ according to 
the rule (\ref{A6}).

The application of $A$ and $A^{\sharp}$ to $\psi_0{}^\flat$
and $(1/\psi_0)^\flat$, respectively, gives
\begin{equation}\label{A5}
	A{\psi_0}^\flat=\frac{1}{\psi_0}, \qquad
	A^\sharp {\left( \frac{1}{\psi_0}\right)}^\flat=-\psi_0\,.
\end{equation}
By virtue of  Eqs.  (\ref{A4})
and
 (\ref{A5}),
the first order operators (\ref{A3}) factorize the shifted Hamiltonian,
\begin{equation}\label{A7}
	H-E_0=A^{\sharp}A\,.
\end{equation}
The alternative product of $A$ and $A^{\sharp}$  gives
the shifted
Darboux (SUSY) partner 
Hamiltonian
$\breve{H}$, 
\begin{equation}\label{A9}
	\breve{H}-E_0=AA^{\sharp}=-\frac{d^2}{dx^2}+\breve{V}(x)\,,
	\quad {\rm with}\qquad
	\breve{V}(x)=V(x)-2\left(\log \psi_0(x))\right)''.
\end{equation}
{The nodes of $\psi_0$ give rise to singularities in $\breve{V}(x)$ not present in $V(x)$. When $\breve{V}(x)$ is required to inherit the regularity of $V(x)$, $\psi_0$ has to be fixed as a nodeless function.}
The intertwining relations
\begin{equation}\label{A10*}
	AH=\breve{H}A\,,\qquad
	A^\sharp\breve{H}=HA^\sharp\,,
\end{equation}
follow 
then
from
(\ref{A7}) and (\ref{A9}).
The operator $\breve{H}-E_0$ annihilates 
the states
${1}/{\psi_0}$ and 
${\left({1}/{\psi_0}\right)^\flat}$. 

In dependence on the nature of $E_0$ and $\psi_0(x)$, 
the Hamiltonians $H$ and $\breve{H}$ are isospectral or almost
isospectral up to eigenvalue $E_0$, which can be present in one system but missing in the other one.
Operators $A$ and $A^\sharp$ realize a mapping between 
two-dimensional eigenspaces of 
the second order differential operators $H$ and $\breve{H}$,
\begin{equation}\label{A11}
	A\psi_E=\breve{\psi}_E\,,\qquad
	A^\sharp\breve{\psi}_E=(E-E_0)\psi_E\,,
\end{equation}
for  any  $E\neq E_0$ and corresponding eigenstates, 
$H\psi_E=E\psi_E$, $\breve{H}\breve{\psi}_E=E\breve{\psi}_E$.

Notice that 
the zero mode ${\psi_0}^\flat$ of $H-E_0$ is mapped by $A$  into the zero mode $1/\psi_0$ of 
$\breve{H}-E_0$ and, analogously, the zero mode ${(1/\psi_0)}^\flat$ of $\breve{H}-E_0$
is mapped by $A^\sharp$ into 
{the}
zero mode $\psi_0$ of $H-E_0$. 
However,  the pre-image  of  
{the}
zero mode  
${\psi_0}^\flat$ of $H-E_0$ 
with respect to the action of $A^{\sharp}$
 is not 
a zero mode of $\breve{H}-E_0$. 
Similarly, the
pre-image of 
{the}
zero mode ${(1/\psi_0)}^\flat$ of $\breve{H}-E_0$
under the action of $A$ does not belong to the kernel of $H-E_0$. 
Instead, the indicated pre-images are the 
\emph{Jordan states}
of the 
corresponding  operators $H-E_0$ and $\breve{H}-E_0$.
Indeed, consider the equation
 \begin{equation}\label{A12}
	A\chi=-{(1/\psi_0)}^\flat\,.
\end{equation}
Its solution is
 \begin{equation}\label{A14b}
	\chi(x)=-\psi_0(x)\left(\int_{x_0}^x\frac{1}{\psi_0(s)}
	({1/\psi_0(s)})^\flat ds+a\right)=-\psi_0(x)\left(\int_{x_0}^x\frac{\alpha+\int_{x_0}^s
	\psi_0^2(r)dr}{\psi_0^2(s)}ds+a\right)\,,
\end{equation}
where $a\in\C$ and $\alpha\in\C$ are
some  constants.
Acting on both sides of (\ref{A12}) by $A^\sharp$,
 one finds that 
 \begin{equation}\label{A15}
	(H-E_0)\chi=\psi_0\,,\quad (H-E_0)^2\chi=0\,.
\end{equation}
This means that  $\chi(x)$ is the Jordan state of $H-E_0$. 
Analogously, we find that 
the solution of equation
$	A^\sharp\lambda={\psi_0}^\flat,$
given by 
$	\lambda(x)=-\frac{1}{\psi_0(x)}\left(\int_{x_0}^x\psi_0(s)
{\psi_0}^\flat(s)ds+b\right)\,,\ 
	b\in\C\,,
$ 
 satisfies 
$	(\breve{H}-E_0)\lambda=
{1}/{\psi_0}\,,
$ 
$	(\breve{H}-E_0)^2\lambda=0$.
Without loss of generality, 
we set
below
$a=0$ in (\ref{A14b}).

Let us take $E\neq E_0$ and consider the
eigenstate equation 
 \begin{equation}\label{A19}
	(H-E)\Psi(x;E)=0\,.
\end{equation}
We suppose that $E$ is sufficiently close to $E_0$ and 
denote $E-E_0=\varepsilon$.  Assuming the analyticity of $\Psi(x;E)$ 
in $E$ in vicinity of $E_0$,
we look for solution of (\ref{A19}) close to 
$\psi_0(x)$  in the form of the Taylor series 
in $\varepsilon$,
 \begin{equation}\label{A20}
	\Psi(x;E)=\psi_0(x)+\sum_{n=1}
	\frac{\varepsilon^n}{n!}\chi_n(x)\,,\qquad
	\chi_n(x)=
	\frac{\partial^n \Psi(x;E)}{\partial E^n}\vert_{E=E_0}.
\end{equation}
Substituting this series into (\ref{A19}), we find that 
$\chi_n(x)$ are defined recursively by 
 \begin{equation}\label{A21}
	(H-E_0)\chi_1(x)=\psi_0(x)\,, 
\end{equation}
 \begin{equation}\label{A22}
	(H-E_0)\chi_n(x)=n\chi_{n-1}(x)\,,\qquad n=2,\ldots .
\end{equation}
Therefore, $\chi_1(x)$ is just the Jordan state that appeared in 
Eq. (\ref{A15}), $\chi_1(x)=\chi(x)$.
The state $\chi_n(x)$ with $n\geq 2$
satisfies relations $(H-E_0)^n\chi_n(x)=\psi_0(x)$,
$(H-E_0)^{n+1}\chi_n(x)=0$, and can be identified
as a higher, $n$-th order Jordan state.

Consider the system $\tilde{H}_\varepsilon=-\frac{d^2}{dx^2}+
V_\varepsilon(x)$ generated from $H$ by applying to the latter 
the Darboux-Crum transformation based on the eigenstates $\psi_0(x)$ and $\Psi(x;E)$,
 \begin{equation}\label{A22b}
	V_\varepsilon(x)=V(x)-2\left(\log W(\psi_0(x),\Psi(x;E))\right)''\,.
\end{equation} 
Taking
 into account 
equation (\ref{A20}),
in  the limit $\varepsilon\rightarrow 0$ we get 
\begin{equation}\label{A22c}
	\tilde{V}:=
	 \lim_{\varepsilon\rightarrow 0} V_\varepsilon (x)=V(x)-2\left(\log W(\psi_0(x),\chi(x))\right)''.
\end{equation}
Relation $W(f,g)=f^2\frac{d}{dx}(\frac{g}{f})$ together with 
Eq. (\ref{A14b}) {gives} us 
\begin{align}
\label{A23}
	\tilde{V}(x)&=V(x)-2\frac{d^2}{dx^2}\log\, \mathcal{I}(x)=
	V(x)-4\frac{\psi_0\psi_0'}{
	\int_{0}^x
	\psi_0^2+\alpha
	}+2\frac{\psi_0^4}{\left(\int_{0}^x
	\psi_0^2+\alpha\right)^2}\,,
\end{align}
where
\begin{equation}\label{caldefI}
	\mathcal{I}(x)= \int_{0}^x\psi_0^2(s)ds+\alpha\,,
\end{equation}
and from now on we set $x_0=0$.

In correspondence with the Darboux-Crum construction,
the Hamiltonian operators $H$, $\breve{H}$ and 
\begin{equation}\label{A24}
	\tilde{H}
	:=
	 -\frac{d^2}{dx^2}+\tilde{V}
\end{equation}
are almost isospectral.
Since in the limit $\varepsilon\rightarrow 0$, 
$A\Psi(x;E_0+\varepsilon)=
A(\psi_0+\varepsilon \chi)=
\varepsilon A\chi=\varepsilon{(1/\psi_0)}^\flat$,
we have 
\begin{equation}\label{A25}
	\breve{H}-E_0=B^\sharp B\,,\qquad
	\tilde{H}-E_0=BB^\sharp \,,
\end{equation}
where 
\begin{equation}\label{A26}
	B=\eta_0\frac{d}{dx}\frac{1}{\eta_0}\,,
	\qquad
	B^\sharp=-\frac{1}{\eta_0} \frac{d}{dx} \eta_0\,,\qquad
	{\rm with}
	\quad
	{\eta_0}:=
	{(1/\psi_0)}^\flat\,.
\end{equation}
There holds  
$B\eta_0=0,\ 
B{\eta_0}^\flat=1/\eta_0\,,
$
$B^\sharp (1/\eta_0)=0,\
B^\sharp {(1/\eta_0)}^\flat=-\eta_0\,,
$
and so, ${\rm ker}\,(\tilde{H}-E_0)={\rm span\,}\{1/\eta_0,{(1/\eta_0})^\flat\}\, $. 
{}From (\ref{A25}) it follows that  
\begin{equation}\label{A28}
	B\breve{H}=\tilde{H}B\,,\qquad
	B^\sharp\tilde{H}=\breve{H}B^\sharp\,.
\end{equation}
The $H$ and $\tilde{H}$ are intertwined then
by the 
second order differential operators,
\begin{equation}\label{A29}
	(BA)H=\tilde{H}(BA)\,,\qquad
	(A^\sharp B^\sharp)\tilde{H}=H(A^\sharp B^\sharp)\,.
\end{equation}
The eigenstates $\psi$ and $\tilde{\psi}$ of $H$ and $\tilde{H}$,
$H{\psi}=E{\psi}$, 
$\tilde{H}\tilde{\psi}=E\tilde{\psi}$,
for $E\neq E_0$
satisfy 
\begin{eqnarray}
 	&&\tilde{\psi}=(BA)\psi=(E_0-E)\psi+
	\frac{\psi_0}{\int^{x}_{0}
	 \psi^2_0(s)ds
	 +\alpha}(\psi\psi_0'-\psi_0\psi'),  \label{tildeu1+}\\ \label{tildeu2}
	&& (A^\sharp B^\sharp)\tilde{\psi}=
	(E-E_0)^2 \psi.
\end{eqnarray}

Notice that in accordance with Eq. (\ref{tildeu1+})
and (\ref{A26}), the eigenstate  $\psi_0{}^\flat$ of $H$ 
of eigenvalue $E_0$
is mapped into 
the corresponding eigenstate of $\tilde{H}$,
\begin{equation}\label{Psi0}
	\widetilde{\psi_0{}\flat}=-\frac{1}{\eta_0}=-
	\frac{\psi_0}{\int^{x}_{0}
	 \psi^2_0(s)ds
	 +\alpha}\,.
\end{equation}
To simplify notations, in what follows we 
re-denote $-\widetilde{\psi_0{}^\flat}$
by $\tilde{\Psi}_0$.

Thus, the 
confluent double step Darboux transformation coincides with the second order 
Darboux-Crum transformation based on the eigenstate $\psi_0(x)$ of 
$H$, $(H-E_0)\psi_0=0$, and the
associated with it Jordan state 
$\chi(x)$ that satisfies $(H-E_0)\chi(x)=\psi_0(x)$, 
$(H-E_0)^2\chi(x)=0$, 
see Eq. (\ref{A22c}). 

The described picture can be generalized further
by applying the confluent Darboux-Crum transformations
based on the eigenstate $\psi_0$ and the 
associated Jordan states 
$\chi_1,\ldots,\chi_n$ of $H$,  see  (\ref{A21}), (\ref{A22}).
A new Hamiltonian ${\tilde{H}}_n$, ${\tilde{H}}_1={\tilde{H}}$,
will be intertwined then with $H$ by the $n$th-order 
operators.
Instead of analyzing such new systems following the line 
presented here, we just notice that 
{a Wronskian 
formulation} 
for
confluent supersymmetric transformation chains  
{was}
 discussed e.g. in \cite{CSUSY4}.

 \section{$PT$-symmetric systems\label{BFsection}}
 
We are going to consider $PT$-symmetric systems with either periodic or 
asymptotically periodic potential. Theoretical aspects of the complex and
 $PT$-symmetric periodic potentials have been 
a
subject of intensive research, see e.g. 
\cite{susypt1,BDM,jones,cervero,shin,samsonovroy}. 
We will focus on the construction of $PT$-symmetric systems
by applying 
the confluent Darboux-Crum 
transformations 
to Hermitian periodic 
Schr\"odinger Hamiltonians. 
We
suppose that the initial system 
is given by
a real, regular,
even
$L$-periodic potential defined on the real line
\begin{equation}\label{B1}
   V(x+L)=V(x), \quad V(-x)=V(x)\,,\quad V^*(x)=V(x)\,. 
\end{equation} 
The
generic properties of the solutions of the equation
\begin{equation}\label{eq1}
 	\left(-\frac{d^2}{dx^2}+V(x)\right)\psi=E\psi\,,\quad E\in\mathbb{R}\,,
\end{equation}
are described then by Floquet's theorem \cite{Megnus}. It tells that the two 
linearly 
independent solutions of (\ref{eq1}), which are neither periodic nor antiperiodic, 
can be written in terms of quasi-periodic functions~\footnote{We call a function 
$f(x)$ quasi-periodic when it satisfies $f(x+L)=c f(x)$ where $c$ is a complex number.}, 
\begin{equation}\label{fs}
	 \psi_{\pm}(x)
	 =e^{\pm ikx}u_{k}(\pm x)\,,\quad u_{k}(x+L)=
	 u_{k}(x)\,, \quad k\in\mathbb{C},\quad \frac{kL}{\pi}\notin\mathbb{Z}\,.
\end{equation}
Here, we have taken into account 
that the spatial reflection operator $P$ is the symmetry of $H$.
We deal with the bounded (``stable'') solutions as long as $k$ is purely real. 
Their energies belong to the interior of the allowed 
bands. 
The functions 
(\ref{fs}) with ${\rm Im}\, k\neq 0$
are unbounded (``unstable'') as they diverge exponentially when $x$ goes towards 
plus or minus infinity. The corresponding energies are non-physical
 and form forbidden bands in the spectrum.
When
  one of the solutions of (\ref{eq1}) is periodic or antiperiodic, 
i.e. when $\frac{kL}{\pi}\in\mathbb{Z}$,
   the second solution can either have the same periodicity or can fail to be 
   quasi-periodic at all~\footnote{When (\ref{eq1}) has a periodic or anti-periodic 
   solution $p(x)$, then the other solution $y(x)$ satisfies $y(x+2L)=y(x)+\gamma p(x)$. 
   The actual value of $\gamma $ depends  on the concrete form of the potential
in   (\ref{eq1})
(one can have $\gamma=0$), see 
\cite{Megnus} for more details.}. The former property corresponds to the solutions
inside the allowed bands, while the latter characterizes the states
at the  edges of the allowed energy bands \footnote{ 
The free particle system is a particular 
example where all physical solutions are periodic. 
Indeed,
the Hamiltonian 
$H =-\frac{d^2}{dx^2}$
is translation invariant
and the solutions $\sin kx$ and $\cos kx$ are
periodic with the period 
$\frac{2\pi}{k}$ 
for any real $k\neq  0$, while the constant solution with $k=0$, 
at the edge of continuous spectrum has
an arbitrary period. 
}. 

In order to get $PT$-symmetric Hamiltonian
 $\tilde{H}$,
 we suppose that the eigenfunction $\psi_0$ 
 of the initial $PT$-symmetric Hamiltonian $H$ is also an eigenstate of the operator $PT$,
\begin{equation}\label{psi0pt}
	PT\psi_0=\epsilon\psi_0\,,
	\quad \epsilon^2=1\,,\quad (H-E_0)\psi_0=0\,,
	\quad E_0\in\mathbb{R}\,.
\end{equation}
The requirement of $PT$-symmetry of $\tilde{H}$ is equivalent to 
$PT\frac{d^2}{dx^2}\log\, \mathcal{I}(x)PT=\frac{d^2}{dx^2}\log\, \mathcal{I}(x)$ {with $\mathcal{I}(x)$ defined in (\ref{caldefI})}.
Taking into account (\ref{psi0pt}) and (\ref{caldefI}), this requirement can be met provided 
that $\alpha$ is purely imaginary, 
\begin{equation}\label{condi}
	 \alpha=i{\alpha}_{{}_R}\,,\quad {\alpha}_{{}_R}\in\mathbb{R}\,.
\end{equation} 
Besides,
 $\alpha_{{}_R}$ has to be (and can be) fixed in a way
 such that $\tilde{V}$ is free of singularities, 
 i.e. $\mathcal{I}(x)$ is nodeless. We suppose this to be the case from now on.

Let us discuss in more detail the consequences of the requirement 
(\ref{psi0pt}) for both stable and unstable solutions. 

First, let us suppose that $\psi_0$ is a linear combination of the stable 
states $g_{\pm}$ with real quasi-momentum, 
\begin{equation}\label{kint}
 	g_{\pm}=a^{\pm} e^{\pm ikx}u_{k}(\pm x)\,,\quad k\in\mathbb{R}\,,
	 \quad k\neq n\frac{\pi}{L},\quad |a^{\pm}|=1\,.
\end{equation}
In definition of $g_-$, we used the fact that $P$ is 
a
symmetry of $H$. 
The phase factors $a^{\pm}$ are fixed such that there holds 
\begin{equation}\label{PTint}
	  PTg_{\pm}=g_{\pm}\,,\quad Pg_{\pm}=g_{\mp}\,,
	  \quad Tg_{\pm}=g_{\mp}\,.
\end{equation}
This implies that any linear combination $b_1g_{+}+b_2g_{-}$ with real $b_1,$ 
$b_2$ is $PT$-symmetric. It is also nodeless provided that $|b_1|\neq |b_2|$; 
this follows from the fact that 
the
real and imaginary parts of the latter linear combination are two independent 
solutions of (\ref{eq1}), and, hence, cannot vanish simultaneously. 

Secondly, we consider the case when $\frac{kL}{\pi}\in\mathbb{Z}$. Let us 
take $\psi_0$ as a linear combination of 
the
states where at least  one of them
  is periodic or antiperiodic and
 $PT$-symmetric. Let
  us denote it $p(x)$. We have 
\begin{equation}\label{fper}
	\psi_0(x)=c_+p(x)+c_-q(x)\,,\qquad p(x+2L)=p(x)\,,                                                                                                                                                                                                                                                                                                                                                                                                         \end{equation}
 where the function $q(x)$ can be written as
$q(x)=p(x) \int_{0}^x
\frac{1}{p(s)^{2}}ds$.  As it was already noted,
such
states are 
associated either with the 
band-edge energies
or  they can correspond to specific energy levels from
 the interior of the allowed energy bands as well.  When $PTp(x)=\epsilon p(x)$, 
 then $PTq(x)=-\epsilon q(x).$ Hence, $\psi_0$ is $PT$-symmetric provided that 
$c_+\in\mathbb{R}$ and $c_-\in i\mathbb{R}$ up to a common multiplicative factor.

Finally, let us consider $\psi_0$ as a linear combination of unstable states (\ref{fs}) 
with complex quasi-momentum,
 $k=k_2-ik_1,$  $k_1> 0$, $k_1,k_2\in\mathbb{R}$. 
One can show that $k_2=\frac{\pi n}{L}$ where $n$ is an integer. 
Indeed, let $f(x)=e^{k_1x}e^{ik_2x}u_k(x)$ 
is an unbounded solution of (\ref{eq1}). 
Then  $f^*(x)$ is also a solution. Considering asymptotic behavior of the two functions, we find that $f(x)=c 
 f^*(x)$ for a constant $c$. 
 We get $c=e^{-2ik_2x}\frac{u_k^*(x)}{u_k(x)}$. Substituting $x\rightarrow x+L$ on the right-hand side, the equation is satisfied provided that $k_2=\frac{\pi n}{L}$. 
 The equation also implies that $u_k(x)$ and $u_k^*(x)$ differ just by a multiplicative constant, and, hence, $u_k(x)$ can be fixed as a real function.
We can write the fundamental solutions in terms of \textit{real} functions  
\begin{equation}
	 f_+=e^{k_1x}u_k(x)\,,\qquad f_-=Pf_+
	 =e^{-k_1x}u_k(-x)\,,
\end{equation}
that satisfy the following relations
\begin{equation}\label{Tf}
 	Tf_{\pm}=f_{\pm}\,,\qquad PTf_{\pm}=f_{\mp}\,.
\end{equation}
To get an eigenstate of $PT$, we have to take the linear combination
\begin{equation}\label{F}
	 F_{\pm}=Cf_+\pm {C}^*f_-\,,\qquad PTF_{\pm}=\pm F_{\pm}\,,
	 \quad \quad C=c_1+ic_2,\qquad c_1,c_2\in\mathbb{R}\,.
\end{equation}
These states are nodeless if $c_1c_2\neq0$. In such a case, the real and imaginary 
parts form the fundamental set of solutions and cannot vanish simultaneously at the same point.

\section{Properties of $\tilde{H}$ \label{PTGENsection}}
In this section, we address the question of square integrability of 
$\tilde{\Psi}_0=-\widetilde{\psi_0{}^\flat}$
and of the asymptotic behavior of $\tilde{V}$. We shall consider separately three 
situations, distinguished by the three qualitatively different forms of $\psi_0$ 
discussed in the previous section. In the first two cases, $\psi_0$ will be 
associated with the energy $E_0$ belonging to  an allowed 
spectral band, whereas in the third case, $E_0$ will be non-physical
eigenvalue of $H$ belonging to the spectral gap. \newline

Let us start with the case where $\psi_0$ is a linear combination of the quasi-periodic functions (\ref{kint}),
\begin{equation}\label{p1}
	 \psi_0(x)=c_+e^{ikx}u_{k}(x)+c_-e^{-ikx}u_k(-x)\,,
\end{equation}
where $c_{\pm}$ are real constants. In order to show that $\tilde{\Psi}_0$
is square integrable, let us suppose that 
$\psi_0$ is a \textit{periodic} function 
with period ${\ell}$, i.e. $\psi_0(x+\ell)=\psi_0(x)$. 
This requirement is satisfied when the periods of $e^{ikx}$ and $u_k$ are 
\textit{commensurable}. It is worth to emphasize that in some cases, ${\ell}$ 
may be reduced to the original period $L$ of the Hamiltonian $H$. 
We have
\begin{eqnarray}\label{w1}
 	\int^{x}_0\psi_0^2(s)ds
	=&
	\int_{0}^{\left[\frac{x}{{\ell}}\right]\ell}\psi_0^2(s)ds
	 +\int^{x}_{\left[
	 \frac{x}{{\ell}}\right]{\ell}}\psi_0^2(s)ds
	 	 ={\left[\frac{x}{
	 {\ell}}\right]}Q_0+\int^{x}_{\left[\frac{x}{{\ell}}\right]
 	{\ell}}\psi_0^2(s)ds\,\,, \label{w2}
\end{eqnarray}
where $Q_0\equiv\int_{0}^{{\ell}}\psi_0^2(s)ds$ and $\left[\frac{x}{{\ell}}\right]$ is the 
integer part of $\frac{x}{{\ell}}$.  
We suppose here that the integral $Q_0$ is nonvanishing. 
Using (\ref{w1}) and boundedness of $\psi_0$, we can see that 
$\tilde{\Psi}_0$
decays asymptotically as 
$1/x$,
\begin{eqnarray}\label{tp1}
 	{\tilde{\Psi}_0} &=&\frac{\psi_0}{\int^{x}_0\psi^2_0(s)ds+
	 i\alpha_{{}_R}}=\frac{\psi_0}{{\left[\frac{x}{{\ell}}
	 \right]}Q_0+\int^{x}_{\left[\frac{x}{{\ell}}\right]{\ell}}
 	\psi_0^2(s)ds+i\alpha_{{}_R}}
	\sim\frac{1}{x}\quad\mbox{for}\quad |x|\rightarrow 
	 \infty\,,
\end{eqnarray}
and, therefore, the eigenstate $\tilde{\Psi}_0$
 is square integrable.

For the second case, let us consider $\psi_0$ as fixed in (\ref{fper}),
\begin{equation}\label{p2}
	 \psi_0=c_+p(x)+c_-q(x)\,,
	\qquad \psi_0(x+2nL)=\psi_0(x)+n\gamma p(x)\,,
	\qquad\gamma\neq 0\,,
\end{equation} 
where $\gamma$ and $c_{\pm}$ are constants. Notice that when $\gamma=0$, 
$\psi_0$ reduces to the form of Eq. (\ref{p1}) and (\ref{w1}) and (\ref{tp1}) apply in this case. 
When $\gamma\neq0$, we can make the same steps that led to (\ref{tp1}), yielding 
\begin{equation}
 	\int_{0}^x\psi_0^2(s)ds=c_0+c_1\left[\frac{x}{2L}\right]+
	c_2\left[\frac{x}{2L}		\right]^2+c_3\left[\frac{x}{2L}
 	\right]^3
	+\int^{x}_{\left[\frac{x}{{2L}}\right]
 	{2L}}\psi_0^2(s)ds\sim x^3\quad \quad\mbox{for}\quad |x|\rightarrow 
 	\infty\,,
\end{equation}
where $c_j$, $j=0,1,2,3$, are constants. 
Hence, the function $\tilde{\Psi}_0$
 is also square integrable  in this case, 
 \begin{equation}
 	\tilde{\Psi}_0=\frac{\psi_0}{\int_0^x\psi_0^2(s)ds+
	 i\alpha_{{}_R}}
	 \sim \frac{1}{x^2}\,,\quad|x|\rightarrow \infty\,.
\end{equation}
In contrast to (\ref{tp1}),  it decays as $1/x^2$.

In the last case,  $\psi_0$ is given as a linear combination of unbounded solutions (\ref{F}). We fix
\begin{equation}\label{p3}
 	\psi_0=F_+=e^{k_1 x}u_{k }(x)+
 	e^{-k_1 x}u_{k}(-x)\,,
\end{equation}
(the analysis for other choice of $\psi_0$ would follow similar steps). 
To make the forthcoming computation 
compact yet easy to follow, let us introduce the following temporal notations, 
\begin{align}
	\tilde{x}\equiv \left[\frac{x}{{L}}\right], \quad 
	\mathcal{J}(x)=e^{2k_1x}u_{k }(x)u_{k }'(x), \quad 
	\mathcal{G}(x)=u_{k }(x)u_{k }(-x),
 	\quad \mathcal{K}(x)=\int_{0}^{x-\tilde{x}{L}} \mathcal{J}(s) ds \,,
 \end{align}
and then we can write
\begin{eqnarray}
	 &&
	 \int_{0}^x\psi_0^2(s)ds
	 =\int_{-x}^x
 	e^{2k_1s}u_{k }^2(s)ds+2\int_0^x u_{k }(s)u_{k }(-s) ds=\nonumber\\
	&&\left[\frac{e^{2k_1s}u_{k }^2(s)}{2k_1}\right]^{x}_{-x}
	-\frac{1}{k_1}\left(\sum_{n=-\tilde{x}}^{\tilde{x}-1}
	e^{2k_1n{L}}\int_{0}^{L} \mathcal{J}(s) 
	ds
	+e^{2\tilde{x}{L}k_1}
	\mathcal{K}(x)
	-e^{-2\tilde{x}{L}k_1}\mathcal{K}(-x)\right)+2
	\int_{0}^x\mathcal{G}(s) ds
	\,=\nonumber\\
	&&\left[\frac{e^{2k_1s}u_{k }^2(s)}{2k_1}
	\right]^{x}_{-x}-\frac{1}{k_1}\left(Q\sinh 2k_1
	\tilde{x}{L}
	+
	e^{2\tilde{x}{L}k_1}\mathcal{K}(x)
	-e^{-2\tilde{x}{L}k_1}\mathcal{K}(-x)\right)+2
	\int_{0}^x \mathcal{G}(s) ds\,,\label{GP}
\end{eqnarray}
where we integrated by parts in the second line and summed over $n$ 
in the third line. Here,
$Q=(\coth k{L} -1)\int_{0}^{{L}}\mathcal{J}(s)ds$
 is a number, whereas $\mathcal{K}(x)=\mathcal{K}(x+L)$
 is a periodic function.  We can see that (\ref{GP}) 
 increases exponentially at large $\vert x\vert$.
  Thus, the eigenstate 
 $\tilde{\Psi}_0$
  decays exponentially for $x\rightarrow\pm\infty$,
\begin{equation}
	 \tilde{\Psi}_0=
	\left\{\begin{array}{cr}\frac{2k_1u_{k }(x)}{u^2_{k }(x)-2\mathcal{K}(x)-Q}e^{-k_1x} \,,&
	\quad x\rightarrow \infty\,,\\
	-\frac{2k_1u_{k }(-x)}{u^2_{k}(-x)-2\mathcal{K}(-x)-Q}e^{k_1x} \,,&
	\quad x\rightarrow -\infty\,,\end{array}\right.
\end{equation}
and represents a quadratically integrable bound state of $\tilde{H}$. 

The Hamiltonian $\tilde{H}$ constructed by fixing $\psi_0$ as either (\ref{p1}) or (\ref{p2}) 
has some remarkable properties. First, we can observe that the potential term $\tilde{V}(x)$
coincides asymptotically with $V(x)$. 
This
can be seen easily from the relation
\begin{equation}\label{tV}
	 \tilde{V}=V-4
	 \tilde{\Psi}_0\psi'_0+2
	 \tilde{\Psi}^2_0\psi_0^2\,.
\end{equation}
For (\ref{p1}), 
where $\psi_0$ is bounded and $\tilde{\Psi}_0$
 decays as 
$1/x$, 
 the function 
$\tilde{V}-V$ disappears as $1/x$. When we have (\ref{p2}), $\psi_0$ has 
linear-like behavior whereas 
$\tilde{\Psi}_0$
decays as $1/x^2$. Hence, the term 
$\tilde{V}-V$ vanishes as $1/x^2$. The periodicity defects $\tilde{V}-V$ are \textit{invisible}; 
apart of the bound state $\tilde{\Psi}_0$, 
an arbitrary eigenstate $\tilde{\psi}$ of $\tilde{H}$ 
coincides asymptotically with an eigenstate $\psi$ of $H$ where $\tilde{\psi}=BA\psi$. 
The wave function does not acquire any phase shift
 when passing through the defect. 
 
These conclusions stem directly from the chain of equalities
\begin{eqnarray}\label{tildeu}
 	  \tilde{\psi}&=&BA\psi=(E_0-E)\psi+\frac{\psi_0}{\int^{x}_0\psi^2_0
 	  +\alpha}(\psi\psi_0'-\psi_0\psi')\\
	&=&(E_0-E)\psi+\tilde{\Psi}_0
	(\psi\psi_0'-\psi_0\psi'),\qquad \tilde{H}\tilde{\psi}=E\tilde{\psi}\,,\label{tildeu2}
\end{eqnarray}
where the second term in (\ref{tildeu2}) vanishes asymptotically ($|\psi_0|$ and $|\psi|$ are bounded), 
implying $\tilde{\psi}|_{x\rightarrow\pm\infty}=(E_0-E)\psi$. 
Let us notice that for $\psi_0=const$ 
(in this case
$H$ corresponds to the free particle), the second term in (\ref{tV}) cancels out.

The invisibility of the periodicity defect in $\tilde{V}$ does not take place when $\psi_0$ 
is fixed as a linear combination of unbounded, exponentially 
growing
states. 
The potential term $\tilde{V}$ of $\tilde{H}$ is asymptotically periodic, however, it does 
not coincide with $V$ in general. We have
\begin{equation}
 	\tilde{V}
	=\left\{\begin{array}{lr}V-8k_1u_k(x)\left(\frac{k_1u_k(x)+
	(u_k(x))'}{u_k^2(x)-Q-2\mathcal{K}(x)}-
	\frac{k_1u_k^3(x)}{(u_k^2(x)-Q-2\mathcal{K}(x))^2}\right),&x		
	\rightarrow \infty\,,\\
	V-8k_1u_k(-x)\left(\frac{k_1u_k(-x)+(u_k(-x))'}{u_k^2(-x)-Q-2
	\mathcal{K}(-x)}-	\frac{k_1u_k^3(-x)}{(u_k^2(-x)-Q-
	2\mathcal{K}(-x))^2}\right)\,,&
	x\rightarrow -\infty\,,
	\end{array}\right.
\end{equation}
which is an ${L}$-periodic and real ($u_k$ is $L$-periodic and real, see (\ref{Tf}) and (\ref{p3})) function. 
Also,
 the function $\tilde{\psi}$ differs asymptotically from $\psi$ as the second term in (\ref{tildeu2}) is not vanishing. 
Notice that when $u_k$ is constant, then $\mathcal{K}(x)=0$, $Q=0$ and $\tilde{V}$ 
coincides asymptotically with $V$. This is the case when $H$ is the Hamiltonian of the free particle.
 
For some specific choices of $H$ and $\psi_0$, the intermediate, $PT$-symmetric 
Hamiltonian $\breve{H}$ can be identified with the original  Hamiltonian $H$ up to a constant displacement of the coordinate, 
\begin{equation}\label{shift}
	\breve{H}(x)=H-2(\psi_0'/\psi_0)'=H(x+c)\,,
\end{equation}
where $c$ preserves~\footnote{Hence, $c$ is either real or its imaginary part coincides
 with imaginary period of $\breve{H}(x)$.}  $PT$-symmetry of $\breve{H}(x)$.
When this is the case, the term $\psi_0'/\psi_0$ has period $L$,  
that
is only possible 
provided
$\psi_0=e^{ikx}u_{k}(x)$. The requirement that $\psi_0$ is an 
eigenfunction of $PT$ forces $k$ to be real. Hence, we can get (\ref{shift}) for 
physical energy $E_0$ and quasi-periodic $\psi_0$ only. The Hamiltonian $\tilde{H}$ 
is intertwined with $\breve{H}$ by 
$B=
\eta_0\frac{d}{dx}\frac{1}{\eta_0}$, 
see (\ref{A26}), 
where $\eta_0$ 
is
a linear combination of $e^{ikx}u_{k}(x+c)$ and $e^{-ikx}u_{k}(-x-c)$. 
To make it satisfy $PT\eta_0=\eta_0$, we take 
\begin{equation}
	 \eta_0=e^{ikx}(c_1u_{k}(x+c)+{c^*_1}PTu_{k}
	 (x+c))+e^{-ikx}(c_2u_{k}(-x-c)+{{c^*_2}}
	 PTu_{k}(-x-c))\,.
\end{equation}
The Hamiltonian $\tilde{H}=H(x+c)-2(\eta'_0/\eta_0)'$ is not necessarily $L$-periodic. 
However, it can have period ${\ell}$
when the periodicities of $e^{ikx}$ and $u_{k}(x)$ are commensurable. 
When this is not the case, $\tilde{V}$ fails to be periodic at all. In the next section, 
we will discuss some examples where these conclusions will be illustrated explicitly.

\section{Examples\label{EXsection}}
\subsection{Reflectionless Systems\label{FP}}
Let us analyze the systems 
generated
by confluent Darboux-Crum
(double-step Darboux-Jordan) transformation 
from the free particle,
\begin{equation}\label{fph}
 	H=-\frac{d^2}{dx^2}\,.
\end{equation}
We will discuss the cases where $\psi_0$ is a linear combination of either stable or unstable solutions of the equation
$(H-E_0)\psi=0$. First, we fix $\psi_0$ as a linear combination of the unstable states corresponding to the energy $E_0=-k_0^2<0$.
Without loss of generality, we can choose
 $\psi_0$ in the following form,
\begin{equation}\label{fpexp}
 	\psi_0=\cosh(k_0x+i\tau)
	 ,\quad k_0>0,\quad \tau\in\mathbb{R}\,.
\end{equation} 
Then, according to
Eqs. (\ref{A23}),  (\ref{A24}) and (\ref{Psi0}),
the  
 explicit form of the Hamiltonian 
 $\tilde{H}$ and of its 
 quadratically integrable bound state 
 $\tilde{\Psi}_0\equiv 
 -\widetilde{\psi_0{}^\flat}$
  can be written as
\begin{eqnarray}\label{trigH}
	 &&\tilde{H}=-\frac{d^2}{dx^2}
	 +2\left(
	\frac{\cosh^4(k_0x+i\tau)}{S_H^2}-
	\frac{k_0\sinh(2(k_0x+i\tau))}{S_H}\right),
	\quad 
	\tilde{\Psi}_0
	=	\frac{\cosh(k_0 x+i\tau)}{S_H}\,,
\end{eqnarray}
where
\begin{eqnarray}
	&&S_H=
	i\tilde{\alpha}+\frac{x}{2}+
	\frac{1}{4k_0}\sinh(2(k_0x+i\tau))\,,
\end{eqnarray}
and  $\tilde{\alpha}=
{\alpha_{{}_R}}
-\frac{1}{4k_0}\sin 2\tau$.
The amplitude of both the potential $\tilde{V}$ and the bound state
 $\tilde{\Psi}_0$
decays 
exponentially for large $|x|$.

As a second case, 
we consider 
$\psi_0$ 
to be a linear combination of stable solutions of $(H-E_0)\psi=0$
with $E_0=k_0^2>0$
by
mixing the plane waves
$e^{ik_0x}$ and $e^{-ik_0x}$,
\begin{equation}\label{fpx-1}
 	\psi_0=\cos(k_0x+i\tau)
	 ,\quad k_0>0,\quad \tau\in \mathbb{R}\,.
\end{equation}
The function 
$\psi_0$
is even with respect to the action of $PT$. 
The explicit form of the Hamiltonian $\tilde{H}$ and of its 
bound state 
$\tilde{\Psi}_0$
is~\footnote{
Some analogous but Hermitian 
system with singularities  on the real line
was considered 
in \cite{MP2}.}
\begin{eqnarray}\label{trigT}
	 &&\tilde{H}=-\frac{d^2}{dx^2}
	 +2\left(
	\frac{\cos^4(k_0x+i\tau)}{S_T^2}+
	\frac{k_0\sin(2(k_0x+i\tau))}{S_T}\right),\quad
	\tilde{\Psi}_0=		
	\frac{\cos(k_0 x+i\tau)}{S_T}\,,\\ 	&&S_T=
	i\tilde{\alpha}+\frac{x}{2}+
	\frac{1}{4k_0}\sin(2(k_0x+i\tau))\,,
\end{eqnarray}
where $\tilde{\alpha}=
{\alpha_{{}_R}}
-\frac{1}{4k_0}\sinh2\tau$.
In contrast with the previous case, the amplitude of both the potential and the bound state
decays now as 	$1/x$  for large $|x|$.

Next, we choose $\psi_0$ as a single stable quasi-periodic eigenfunction of 
$H$,
\begin{equation}\label{fpd}
 \psi_0=e^{ik_0x}\,.
\end{equation}
In this case, $\tilde{H}$ can be written as
\begin{eqnarray}\label{fpph}
	 \tilde{H}=-\frac{d^2}{dx^2}+\frac{2k_0^2}{\sin^2(k_0 x+i\tau)}\,,
	 \qquad \tau=\frac{1}{2}\log (1+2k_0 
	{\alpha_{{}_R}}
	 )\,,
\end{eqnarray}
which 
corresponds to the $PT$-regularized trigonometric P\"oschl-Teller 
potential  
{\cite{ZnojilPT}~\footnote{For a 
wider
class of such 
systems see also \cite{MFPT}.}. 
For this choice of $\psi_0$, the intermediate Hamiltonian $\breve{H}$ reduces to $H$. 
Hence, $H$ and $\tilde{H}$ 
are effectively intertwined by 
the
first-order 
Darboux
transformation.
It is worth noticing that the periodic 
Hamiltonian $\tilde{H}$ has a non-degenerate energy level at $E=E_0$. 
Indeed, one of the solutions of $(\tilde{H}-E_0)f=0$
given by  (\ref{Psi0}) is a 
bounded
periodic function 
	$\tilde{\Psi}_0=-\widetilde{\psi_0{}^\flat}$,
\begin{equation}
	\tilde{\Psi}_0=
	\frac{1}{\sin(k_0 x+i\tau)}\,.
\end{equation}
The second solution 
$
\tilde{\Psi}_{0}{}^\flat$
grows asymptotically as $\vert x\vert$ and, hence, it fails to be stable quasi-periodic state.

Finally, let us 
take $\psi_0$ 
  as a linear combination of the band-edge states corresponding to $E_0=0$, i.e.
\begin{equation}\label{fpx-2}
 	\psi_0=bx+ic\,,
\end{equation}
where $b\neq 0$ and $c$ are real parameters. Notice that (\ref{fpx-2}) can be obtained from (\ref{fpexp}) or (\ref{fpx-1}) in the limit $k_0\rightarrow 0$.
As we found in the preceding section, 
in this case
the potential $\tilde{V}$ has to decay asymptotically as 
$1/x^2$.
The actual form of $\tilde{V}$ and of the bound 
(quadratically integrable)
state $\tilde{\Psi}_0$ 
is given by
\begin{eqnarray}\label{rat}
 	&&\tilde{H}=-\frac{d^2}{dx^2}-\frac{2(bx+ic)}{S_C}\left(2b-\frac{(bx+ic)^3}{S_C}\right)\,,
 	\quad 
	\tilde{\Psi}_0=
	\frac{bx+ic}{S_C}\,,\\ &&S_C=icx\,(bx+ic)+\frac{b^2x^3}{3}+
 	i
	{\alpha_{{}_R}}\,.
\end{eqnarray}
For $c=0$, we get the potential 
whose real singular analog
was 
discussed e.g. in \cite{drazin} in the context 
of solutions of the
KdV equation. For $b=0$, we get a regularized two-body Calogero potential
\begin{equation}
 	\tilde{V}=\frac{2c^4}{(
	c^2x-i
	{\alpha_{{}_R})^2}}\,.
\end{equation}
Let us mention that a similar system, the $PT$-regularized Calogero model with both the 
centrifugal barrier and confining $x^2$ potential term, 
was considered
  in 
\cite{ZnojilCalogero1}, \cite{ZnojilCalogero2}, where its spectral properties were discussed in detail.

The fact that $\tilde{H}$ is intertwined with the free particle by 
the confluent
 Darboux-Crum
 transformation implies that the the systems described by $\tilde{H}$
  are \textit{reflectionless} (the known reflectionless systems are related with the free 
  particle system by 
the 
  Darboux-Crum transformation, see 
  e.g.
   \cite{mateev}).
As we discussed in the preceding section, 
the systems constructed by fixing $\psi_0$ 
as (\ref{fpx-1}) or (\ref{fpx-2}) possess 
\emph{invisible potential barriers} since
there is no phase-shift of wave functions.  

When $\psi_0$ is a linear combination of  
unbounded states (\ref{fpexp}), then the potential $\tilde{V}$ is still reflectionless.  
However, when we substitute (\ref{fpexp}) and (\ref{trigH}) into 
(\ref{tildeu}),  
(\ref{tildeu2}),
we get the following asymptotic form of the wave 
functions
in continuous part of the spectrum,
\begin{equation}
	 \lim_{x\rightarrow\infty}BAe^{\pm ikx}=\left(E_0-E+
	 2k_0(k_0\mp ik)\right)e^{\pm ikx}=(k_0\mp i k)^2e^{\pm ikx}\,,
\end{equation}
\begin{equation}
 	\lim_{x\rightarrow-\infty}BAe^{\pm ikx}=
\left(E_0-E+2k_0(k_0 \pm ik)\right)	e^{\pm ikx}=(k_0\pm i k)^2e^{\pm ikx}\,.
\end{equation}
Hence, the functions acquire nontrivial phase shifts
by the interaction  with the potential, that makes
the  potential detectable in principle.

To illustrate our results, 
let
us consider propagation of a light beam in the optical crystal with the refractive index described by $\tilde{V}$. For the reflectionless systems discussed in this section, one can find explicit solutions of the Helmholtz equation (\ref{sit1}). They can be obtained with the use of the intertwining operator $BA$ and the solution of the ``time"-dependent 
free
Schr\"odinger equation 
$i\partial_z{\psi}(x,z)=-\partial_x^2{\psi}(x,z)$ that has the following form 
 \begin{align}\label{freepsi}
	\psi(x,z)=\frac{\sigma}{\sqrt{\sigma^2+ 2iz }} 
	e^{i \omega_0 (x-x_0-\frac{1}{2} v_0 z)}e^{\textstyle -	
	\frac{(x-x_0-v_0z)^2}{2 \left(\sigma^2+2 i z\right)}}\,.
\end{align}
For $z=0$, it coincides with the Gaussian wave packet $\psi(x,0)=e^{i \omega_0 (x-x_0)}e^{ -(x-x_0)^2/2\sigma^2}$. Here, $x_0$ is the initial position of the wavepacket, $\sigma$ is the width parameter while $\omega_0$ and $v_0=2 \omega_0$ are the 
analogues 
of 
{the 
wave number and 
and group velocity}, 
respectively.
Then the solution of the equation (\ref{sit1}) with $\tilde{V}$ can be constructed as follows (notice that the intertwining operator $BA$ is $z-$independent)
\begin{align}\label{full1}
	 i\partial_z\tilde{\psi}(x,z)=[-\partial_x^2+\tilde{V}(x)]\tilde{\psi}(x,z), 
	 \qquad  \tilde{\psi}(x,z)=BA\psi(x,z) \,.
\end{align}
The intensity of the wave packet $|\tilde{\psi}(x,z)|^2$ is depicted in the Fig. \ref{plots}, 
where the absence of any reflections
on the potential barrier is manifest. 
In order to see the contrast with a 
reflective potential barrier, 
we plot in  Fig. \ref{plots2} the intensity of the wave 
packet  propagating in the potential given by 
only the real part of a complex reflectionless potential.
\begin{figure}[h!]
	\centering
	\subfloat[ 
	]
	{ \includegraphics[scale=0.65]{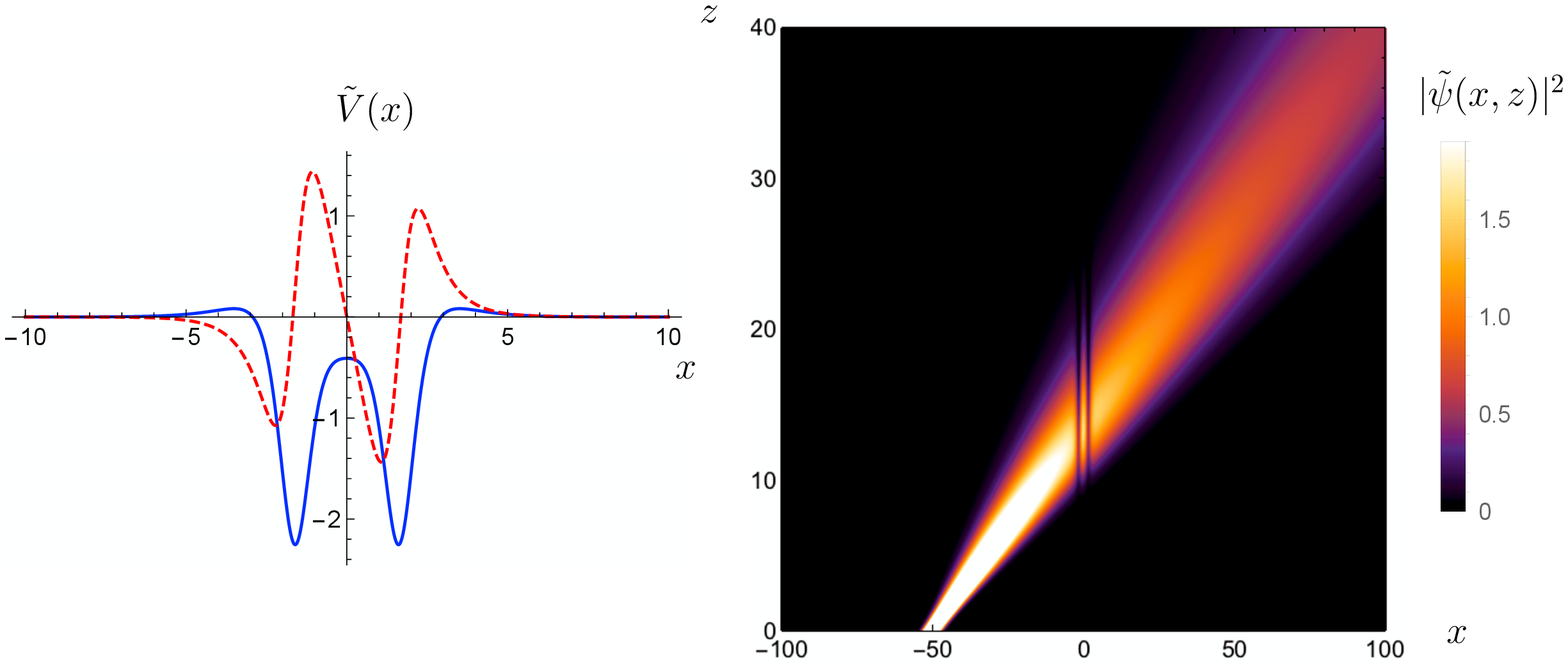}}\\
	\subfloat[]
	 {\,\, \includegraphics[scale=0.65]{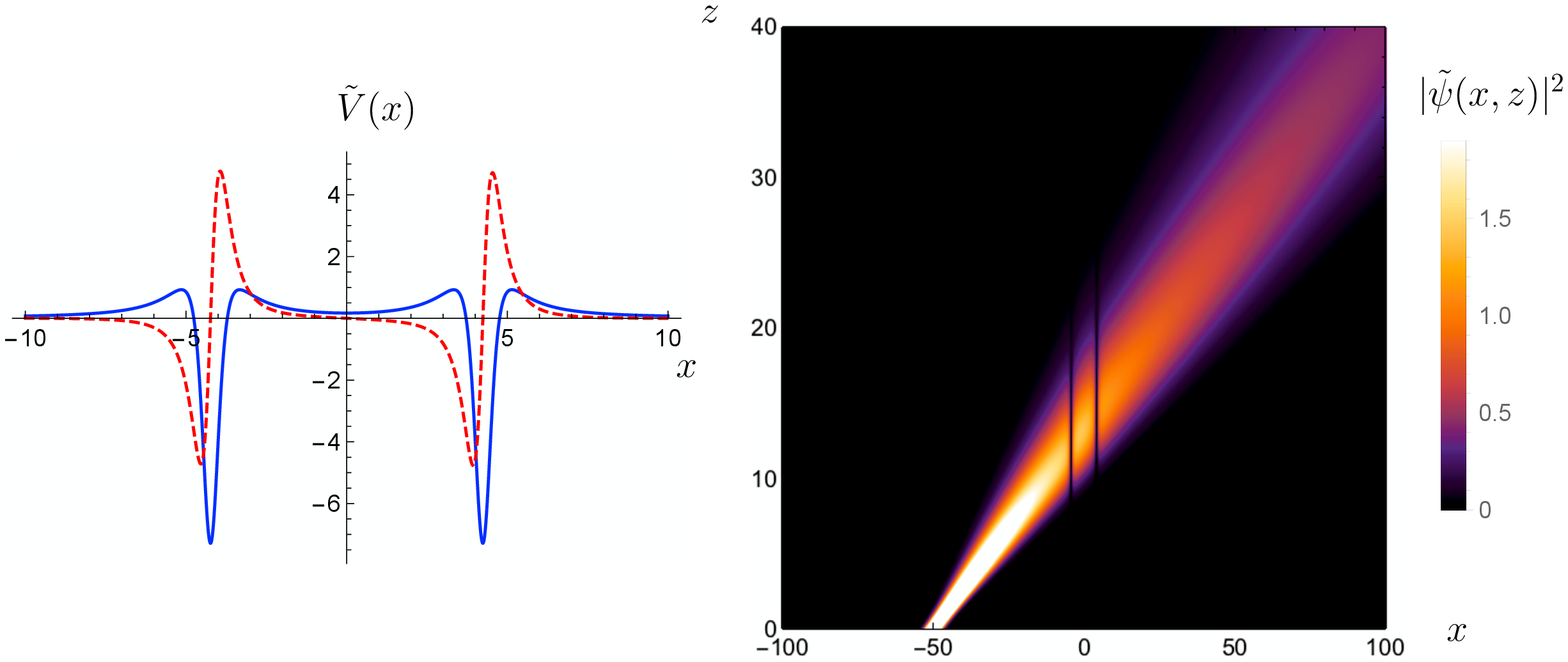}}\\
	\subfloat[
	]
	{\,\, \includegraphics[scale=0.65]{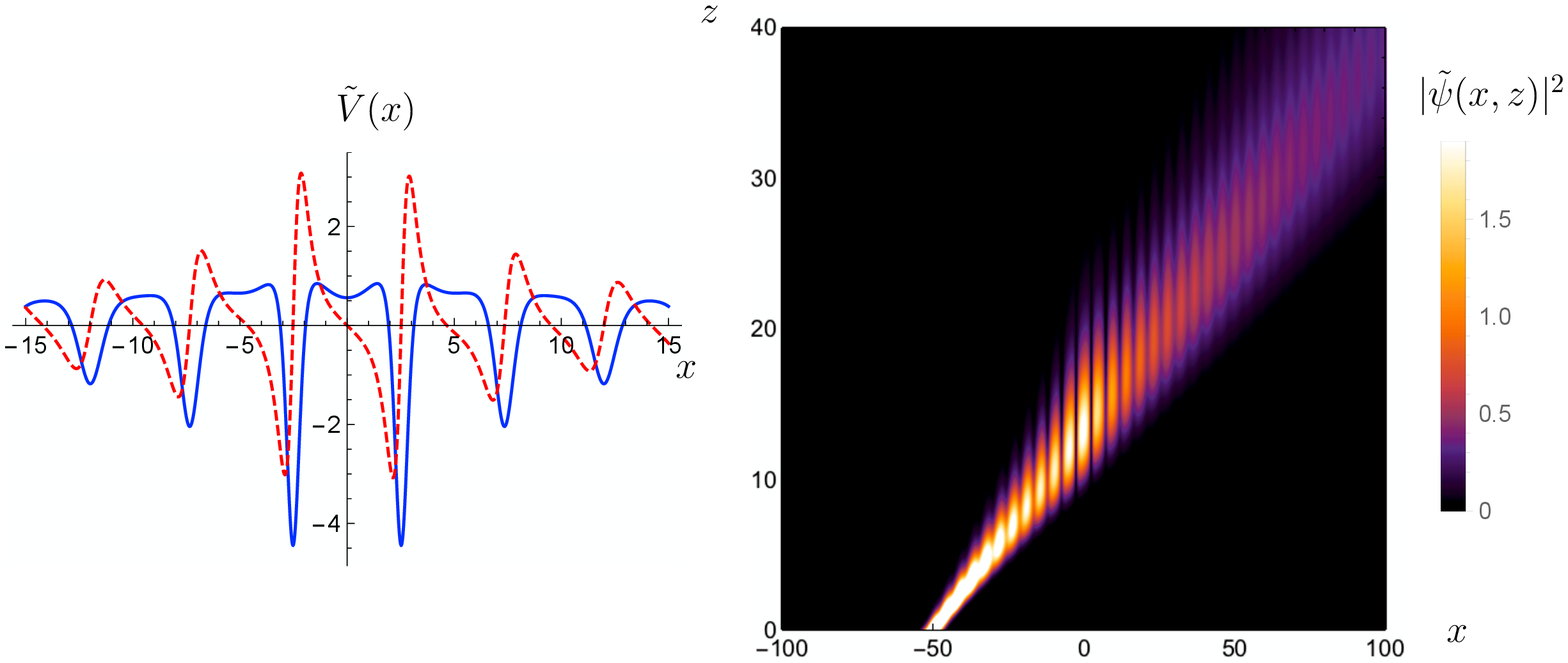}}
\label{plots}
\caption*{} 
\phantomcaption
\end{figure}
\clearpage
\begin{figure}[h!]
 \ContinuedFloat
	\centering
	\subfloat[]
	{\,\, \includegraphics[scale=0.65]{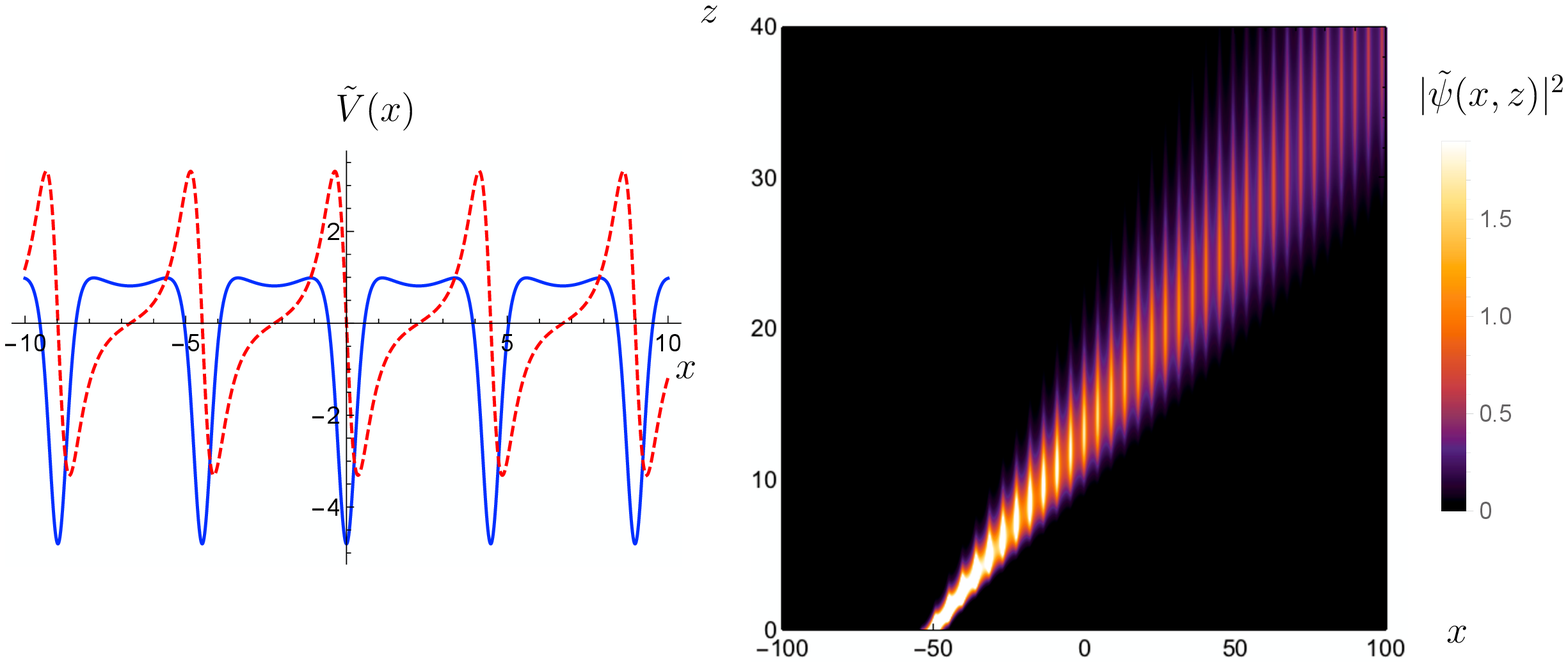}}
	\caption{Reflectionless optical
	{potentials}. 
On the left:	
	$PT$-symmetric potential $\tilde{V}(x)=
	-2\frac{d^2}{dx^2}\log\, \left(\int_{0}^x\psi_0^2(s)ds+
	i \alpha_{{}_R} \right)$
	discussed in section \ref{FP}. 
	The solid blue (dashed red) 
	line represents the real (imaginary) part. 
	On the right:	
	intensity $|\tilde{\psi}(x,z)|^2$ of the light beam  when 
	propagating through the optical 
	{potential}. 
	There is a bound state of energy 
	 with 
	 (a)
	  $E_0=-k_0^2$, 
	  {(b)} $E_0=0$, 
	  {(c)}~$E_0=k_0^2$. 
	The periodic system 
	(d)
	 has no bound states. The respective $\psi_0$ and parameters for the construction of the potentials in each case are given as follows. (a) $\psi_0=\cosh(k_0x+i\tau)$,  	${\alpha_{{}_R}}=1.15$, $k_0=0.7$, $\tau=1.4$ (b) $\psi_0=b x +i c$, ${\alpha_{{}_R}}=-2.8$, $b=-0.26$, $c=-0.5$ (c) $\psi_0=\cos(k_0x+i\tau)$, 	${\alpha_{{}_R}}=4$, $k_0=0.7$, $\tau=1$ (d) $\psi_0=e^{i k_0x}$,  ${\alpha_{{}_R}}=1$, $k_0=0.7$. In $\tilde{\psi}(x,z)$, we fix
        $\sigma=2$, $x_0=-50$ and $\omega_0=\pi/2$ in 
        all 
        the
        cases, see (\ref{freepsi}) and (\ref{full1}).  }
        \label{plots}
\end{figure}


\begin{figure}[h!]
\centering
\includegraphics[scale=0.35]{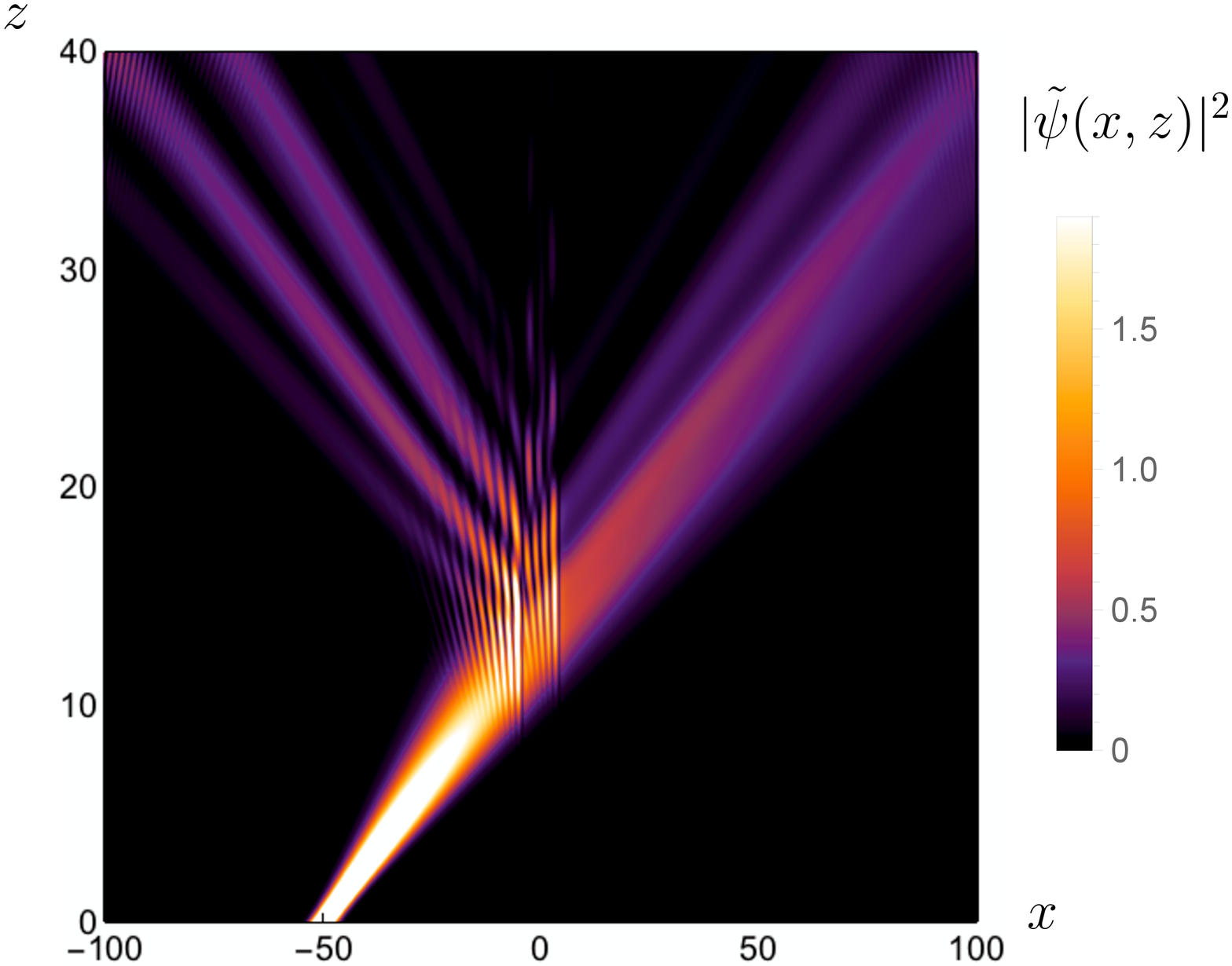}
\caption{
Reflective
optical potential. Intensity $|\tilde{\psi}(x,z)|^2$ of the light beam  when 
	propagating through the \emph{real} part of the potential, ${\rm Re}
	 [\tilde{V}(x)]$, in the model (\ref{rat}).  
	 The parameters are the same as in (b) of Fig. \ref{plots}. 
	The two main 
	reflections
	 that appear about
	  $x=0$ correspond to
	   those
	   produced by each potential well, see (b), Fig. \ref{plots}.}
	 \label{plots2}
\end{figure}

\newpage

\subsection{One-gap systems\label{LAMEsection}}
Let us take
now the initial  
system described by
the one-gap Lam\'e Hamiltonian,
\begin{equation}\label{lameh}
    H=-\frac{d^2}{dx^2}+V(x)\,,\qquad
    V(x)= 2\kappa^2 {\rm sn}^2\,
    (x,\kappa)-\kappa^2=-2\, {\rm dn}^2\, (x,\kappa)+1+\kappa'^2\,.
\end{equation}
The operator has period $2K$,
 where $K\equiv K(\kappa)$ is the complete elliptic integral of the first kind, 
 $K(\kappa)=\int_0^{\pi/2}(1+\kappa^2\sin^2\phi)^{-1/2}d\phi$, 
$\kappa\in(0,1)$ is  the modular parameter, and
 $\kappa'=\sqrt{1-\kappa^2}$ is the complementary 
 modular 
 parameter~\footnote{
Lam\'e potential 
(\ref{lameh})
possesses also a second, purely imaginary period
$2iK'$, $K'=K(\kappa')$, 
that is behind its one-gap nature.}.
The spectrum of the 
Schr\"odinger operator
 $H$ consists of two bands. The finite 
(valence) band is formed by 
the energies $E\in [0,  \kappa'{}^2]$,
and the infinite (conduction)
band stretches over the interval $E\in [1,\infty)$. The functions 
\begin{equation}\label{lamef}
    \Psi_\pm^{a}\left(x\right)=
    \frac{{\rm H}\left(x\pm a\right)}{\Theta\left(x\right)}
    \exp\left[
    \mp x {\rm Z}\left(a\right)\right]\,
\end{equation}
solve the eigenstate problem
\begin{equation}
 	H\Psi_\pm^{a}\left(x\right)=E(a)\Psi_\pm^{a}\left(x\right),\quad E(a)={\rm dn}^2\, 
	 (a,\kappa)\,.
\end{equation}
Here $\Theta$, ${\rm H}$,  and ${\rm Z}$ are
Jacobi's  Theta, Eta and Zeta functions, while
parameter $a$ can take arbitrary complex 
values. The functions ${\rm dn}\, (x,\kappa)$, 
as well as ${\rm cn}\,(x,\kappa)$ and ${\rm sn}\,(x,\kappa)$ 
that will be used later on, are 
Jacobi elliptic functions.

The actual value of $a$ determines whether the function is bounded or unbounded. 
Let us write
$a=\beta+i\gamma$, $\beta,\gamma\in\R$. The wave functions are unbounded (non-physical)
when the corresponding energies belong to the lower infinite band, $E\in (-\infty,0)$,
or to the finite spectral gap, $E\in (k'{}^2,1)$. 
This happens provided  $\beta\in (0,K)$,
 and $\gamma=K'$ or $\gamma=0$, respectively. The two bands of physical eigenvalues 
correspond to
 $\gamma=[0,K']$, 
and $\beta=K$  for a finite (valence), 
 and  $\beta=0$ for  infinite (conduction) bands.
The properties of the functions (\ref{lamef}) with respect to 
the parity 
and complex
conjugation were discussed in detail in 
  \cite{longwork},
  and their behavior under the composite 
  $PT$ transformation is given by
\begin{eqnarray}
	PT\Psi_\pm^{iK+\beta}\left(x\right)&=&
	e^{i\frac{\beta\pi}{K}}\Psi_\mp^{iK+\beta}\left(x\right)
	\qquad
	 \mbox{for}\quad  \beta\in(0,K),	\label{f1}\\
	PT\Psi_\pm^{\beta}\left(x\right)&=&-
	\Psi_\mp^{\beta}\left(x\right)\qquad \mbox{for}\quad \beta	
	\in(0,K),\label{f2}\\
	PT\Psi_\pm^{K+i\gamma}\left(x\right)&=&
	\Psi_\pm^{K+i\gamma}\left(x\right) 
	\qquad \mbox{for}\quad \gamma\in[0,K'],\label{f3}\\
	PT\Psi_\pm^{i\gamma}\left(x\right)&=&-
	\Psi_\pm^{i\gamma}\left(x\right) \qquad 
	\mbox{for}\quad \gamma\in[0,K'],\label{f4}
\end{eqnarray}
where (\ref{f1}) and (\ref{f2}) correspond 
to the lower semi-infinite forbidden band 
and the finite spectral gap, while 
Eqs. (\ref{f3}) and (\ref{f4}) correspond to the finite and infinite allowed bands, 
respectively.

As discussed in the preceding section, 
we can construct operators $\tilde{H}$ with bound states of arbitrary energy 
by taking
$\psi_0$ as an appropriate
 linear combination of (\ref{lamef}). In the current case, 
we deal with the situation where analytical calculation of $\tilde{V}$ 
and
 $\tilde{\psi}_0$ 
gets exceedingly difficult in general due to the term  $\int_0^x\psi_0^2(s)ds$ whose 
analytical form is rather unreachable. We focus here to the special cases 
where $\psi_0$ is 
$2K$-periodic (${\rm dn}\,(x,\kappa)$), 
or
 $4K$-periodic (${\rm cn}\,(x,\kappa)$ and ${\rm sn}\,(x,\kappa)$),
and corresponds to the band-edge energies. 
The band edge states  
and eigenvalues
are given by the relations
\begin{equation}
 	H \,{\rm dn}\,(x,\kappa)=0,
	 \quad H \,{\rm cn}\,(x,\kappa)=k'{}^2\cdot
	 {\rm cn}\,(x,\kappa)\,,
	 \quad H \,{\rm sn}\,(x,\kappa)=1\cdot{\rm sn}\,(x,\kappa)\,.
\end{equation}
Taking
$\psi_0$ as one of the band-edge states,
we get
\begin{eqnarray}
	\tilde{H}=H+\frac{4\kappa^2 \,{\rm sn}\,(x,\kappa) \,{\rm cn}\,(x,\kappa)
 	\,{\rm dn}\,(x,\kappa) }{S_E}+2\frac{F_E}{S_E^2}\,,
\end{eqnarray}
 where we abbreviated 
\begin{eqnarray} 
	F_E=\left\{\begin{array}{r}
	\kappa^4\,{\rm sn}^4(x,\kappa)\,,\\
	\kappa^4\,{\rm cn}^4(x,\kappa)\,,\\
	\,{\rm dn}^4(x,\kappa)\,,
\end{array}\right.
\quad
	S_E=\left\{\begin{array}{rr}-x-i\kappa^2\tilde{\alpha}+{\rm E} (x,\kappa),&\psi_0=\,{\rm sn}\,(x
	,\kappa)\,,\\
	-(1-\kappa^2)x+i\kappa^2\tilde{\alpha}+{\rm E} (x,\kappa),
	&\psi_0=\,{\rm cn}\,(x
	,\kappa)\,,
	\\
	i\tilde{\alpha}+{\rm E} (x,\kappa),&\psi_0=\,{\rm dn}\,(x
	,\kappa)\,.
	\end{array}\right.
\end{eqnarray}
Here, ${\rm E} (x,\kappa)$ is the incomplete elliptic integral
of the second kind, ${\rm E} (x,\kappa)=\int_0^x {\rm dn}^2 (s,\kappa)\, ds $.
The extra potential terms (in addition to initial $V(x)$
from (\ref{lameh}))
vanish
for large $|x|$
 (due to asymptotically linear behavior of $S_E$),
so that $\tilde{V}$ coincides asymptotically with $V$. 
In \cite{longwork}, it was found that real 
periodicity defects, 
associated with discrete energies in the 
forbidden bands,
induce phase shifts
  of the wave functions. In the current case, 
a  $PT$-symmetric defect 
associated with the band-edge energy does not alter 
the asymptotic form of the wave functions.

Despite the 
explicit
analytical form of $\tilde{H}$ for generic choice of $\psi_0$ is 
unreachable, we can still make some interesting conclusions.
Let us discuss briefly the properties of the intermediate Hamiltonian 
$\breve{H}$ when $\psi_0=\Psi_{\pm}^{a}(x)$.  
As it was
discussed in  \cite{longwork}, in this case 
 $\breve{H}$ coincides with the original but displaced Hamiltonian $H$,
\begin{equation}\label{ld}
 	\breve{H}(x)=H(x+a+iK')-E_0\,,
\end{equation}
where $K'=
K(\kappa')$. Notice that $H(x+a+iK')$ is $PT$ symmetric 
 whenever $a=i\gamma$ or $a=K+i\gamma$ for $\gamma=[0,K']$, i.e. when 
 $\Psi_{\pm}^{a}(x)$ corresponds to the physical state. We can fix the solutions 
 of $(\breve{H}-E_0)f=0$ in the following form,
\begin{equation}\label{ad}
	 \psi_\pm=\Psi_{\epsilon}^{a}(x+a+iK')\pm 
	 PT\Psi_{\epsilon}^{a}(x+a+iK'),\quad PT\psi_{\pm}=		
	 \pm\psi_{\pm}, 
\end{equation}
where $\epsilon$ equals either 
$+$ or $-$.
By the construction, (\ref{ad}) 
are eigenfunctions of $PT$. They are also quasi-periodic and bounded. 
In order to construct the Hamiltonian $\tilde{H}$, we fix $B$ in terms of  $\eta_0$ 
which is a linear 
combination of $\psi_+$ and $\psi_-$.
However, taking into account that 
\begin{align}
	PT\Psi_{\pm}^{a}(x+a+iK')=\left\{\begin{array}{l}
	-\Psi_{\pm}^{a}(x+a+iK'), \quad a=i\gamma\,,\\
	e^{\pm2iK(\frac{\pi}{2K}-i\, {\rm Z}(a))}\Psi_{\pm}^{a}(x+a+iK'), 
	\quad a=K+i\gamma\,,
	\end{array}\right.
\end{align}
one can check  that it is sufficient
to consider only $\eta_0=\Psi_{\pm}^{a}(x+a+iK')$. 
This
gives
\begin{align}
	\tilde{H}=\breve{H}-2(\log\eta_0)''=H(x+a+iK')-2(\log\eta_0)''=\left\{\begin{array}{l}
	 H(x+2i\gamma), \quad \eta_0=\Psi_{+}^{a}(x+a+iK')\,,\\
	  H(x), \quad \eta_0=\Psi_{-}^{a}(x+a+iK')\,.
\end{array}\right.
\end{align}
Clearly, $\tilde{H}$ has no bound state as 
$1/\eta_0$ is not quadratically integrable.
Darboux transformation based on $\eta_0$ produces from $\breve{H}$ 
either the initial Hamiltonian $H(x)$ or its complex shifted copy $H(x+2i\gamma)$.

Finally, let us make a brief comment on the 
unbounded states (\ref{lamef}) that can be written as
 $f_{\pm}=e^{k_1x}e^{ik_2x}u_{k_0}(x)$, $k_1,k_2\in \mathbb{R}$. 
 In the one-gap system, the value of $k_2$ distinguishes the valence 
 and conduction bands; it is equal to $0$ for the states from the lower 
 forbidden band, 
whereas $k_2=\frac{\pi}{2K}$ for the finite gap, 
see \cite{longwork} for  the  details.

\section{
Hidden and exotic supersymmetric structures associated with
 finite-gap system 
 $\tilde{H}$,
 and Jordan states
\label{hifi}}
Spectrum of a generic periodic Hamiltonian possesses an infinite number of gaps 
and bands. However, there is a class of systems whose spectra contain just a finite 
number of gaps. If we denote the number of gaps as $n$, the free-particle 
Hamiltonian (\ref{fph}) represents the system with $n=0$, whereas the one-gap
 Lame Hamiltonian (\ref{lameh}) qualifies for $n=1$. The peculiarity of having 
 a finite number of gaps in the spectrum is associated with the existence of an integral 
 of motion known as Lax-Novikov operator $\mathcal{L}$. This integral is 
 represented by a 
 $(2n+1)$-order differential operator. 

 In correspondence with the Burchnall-Chaundy theorem
 \cite{BC1,BC2,Ince,Krich},
 its square is a polynomial in $H$,
\begin{equation}
 	\mathcal{L}^2=\prod_{m=1}^{2n+1}(H-E_m)\,,
	\qquad [H,\mathcal{L}]=0\,.
\end{equation}
Here, $E_m$ are the energies corresponding to the edges of the allowed energy bands. 

The Lax-Novikov integral reflects spectral degeneracy of the finite-gap system. 
It distinguishes 
two states corresponding to the doubly-degenerate energies 
in the interior of the bands. 
Particularly, in the one-gap case,
the Bloch states (\ref{lamef}) of the same energy 
are eigenstates of  $\mathcal{L}$ of eigenvalues
differing in sign.
For the free particle (\ref{fph}), the Lax-Novikov integral coincides with the 
momentum operator $\mathcal{L}=-i \frac{d}{dx}$. It annihilates a constant, 
which is the wave 
function
 corresponding to zero energy. In 
the
 case of  the
 one-gap Lam\'e 
system, the integral acquires a
rather nontrivial form of the third order differential 
operator, that annihilates the band edge states ${\rm sn}(x,\kappa)$, 
${\rm cn}(x,\kappa)$ and 
${\rm dn}(x,\kappa)$,
see \cite{trisusy1,trisusy2,longwork} 
for its explicit form.

The operator $\tilde{H}$ constructed from a finite-gap Hamiltonian $H$ 
possesses the integral of motion obtained by a 
dressing 
procedure,
\begin{equation}
	 \tilde{\mathcal{L}}=BA\mathcal{L}A^{\sharp}B^{\sharp}\,,
	 \qquad [\tilde{H},\tilde{\mathcal{L}}]=0\,.
\end{equation}
When $\mathcal{L}$ is of order $2n+1$, then $\tilde{\mathcal{L}}$ has order $2n+5$.
Making use of
the intertwining relations, we 
 find that
$ \tilde{\mathcal{L}}$
satisfies the relation
\begin{equation}\label{tilLdH}
	 \tilde{\mathcal{L}}^2=(\tilde{H}-E_0)^4\prod_{m=1}^{2n+1}(\tilde{H}-E_m)
	 \,.
\end{equation}
The operator $\tilde{\mathcal{L}}$ inherits the properties of $\mathcal{L}$.
Bloch solutions 
of the equation $(\tilde{H}-E)\tilde{\psi}=0$ 
corresponding to the 
double-degenerate energy level $E$
are also eigenstates of $\tilde{\mathcal{L}}$
of eigenvalues differing in sign.
When 
$E$ corresponds to one of the non-degenerate band-edge energies 
$E_m$, $m=1,...,2n+1$, the corresponding physical states are annihilated by 
this operator. Let us clarify the point where $\mathcal{L}$ and 
$\tilde{\mathcal{L}}$ differ from each other. 
For the purpose we
 shall analyze the kernel of $\tilde{\mathcal{L}}$. 
 
We suppose that $E_m\neq E_0$ for $m=1,...,2n+1$. There are $2n+1$ 
states $\tilde{\psi}_m$ in the kernel 
which
are images of the states $\psi_m$ 
annihilated by $\mathcal{L}$,
\begin{equation}
	 \tilde{\psi}_m=BA\psi_m\,,\quad \mathcal{L}\psi_m=0\,,
	 \quad m=1,...,2n+1\,.
\end{equation}
Let us define auxiliary functions $\Sigma_2$ and $\Sigma_3$ that satisfy 
$ \mathcal{L}\Sigma_2=\psi_0$ and $\mathcal{L}\Sigma_3=\chi.$ Then we 
find
the remaining four vectors from the kernel of $\tilde{\mathcal{L}}$,                                                                                 
\begin{eqnarray}
	&& f_0=\eta_0^{-1}\,,\qquad f_1=\eta_0^{-1}\int_{0}^x\frac{\eta(s)}{\psi_0
	(s)}ds\,,\\
	&& f_2=\eta_0^{-1}\int_{0}^x\frac{\eta_0(s)}{
	\psi_0(s)}\int_0^s \psi_0(r) \Sigma_2(r)dr ds\,,
	\qquad 	f_3=
	\eta_0^{-1}\int_{0}^x\frac{\eta_0(s)}{\psi_0(s)}\int_0^s \psi_0(r) \Sigma_3(r)
	drds\,.
\end{eqnarray}
Using
 the explicit form (\ref{A3}) and (\ref{A26}) of $A^{\sharp}$ and $B^\sharp$, 
one can check that there holds
\begin{equation}
 	B^{\sharp}f_1=-\psi_0^{-1}\,,
	\qquad 
	A^\sharp B^\sharp f_2=\Sigma_2\,,
	\qquad  A^\sharp B^\sharp 	f_3=\Sigma_3\,.
\end{equation}
The functions $f_1$,  $f_2$ and $f_3$ are
the
Jordan states of $\tilde{H}$ associated with $E_0$,
\begin{equation}
	 (\tilde{H}-E_0)^jf_j=\frac{1}{\eta_0}\,,\quad
	  (\tilde{H}-E_0)^{j+1}f_j=0\,, \qquad j=1,2,3\,.
\end{equation}

As it was shown in
\cite{trisusy1,trisusy2},
 the finite-gap systems possess a hidden $N=2$ 
superalgebra graded by
 the parity operator $P$. 
The two supercharges, anticommuting with $P$, are given 
by
 $\mathcal{L}$ and $iP\mathcal{L}$. 
It was showed there that the superalgebraic structure is preserved by the 
Darboux-Crum transformation based on the band-edge states. In our current case, the
 intertwining operators $BA$ and $A^{\sharp}B^{\sharp}$ are $PT$-symmetric, however,
  they do not need to have a definite parity with respect to 
  the space inversion 
  $P$. Hence, as neither 
  $\tilde{H}$ nor $\tilde{\mathcal{L}}$ have definite parity, $P$ is prevented from being a 
  viable grading operator. Instead, we can consider the $PT$ symmetry operator, which 
  commutes both with $H$ and $\mathcal{L}$. Let us define 
  $\mathcal{L}_1=i\mathcal{L}$ 
  and 
  $\mathcal{L}_2=iPT\mathcal{L}$.
   These operators 
     anticommute with $PT$, and generate
     the $N=2$ supersymmetry of $H$, 
\begin{equation}\label{susyH}
	 [\mathcal{L}_a,H]=0\,,\quad \{\mathcal{L}_a\,,
	 \mathcal{L}_b\}=-\delta_{ab}\prod_{n=1}^{2n+1}(H-E_n)\,,	
	 \quad a,b=1,2\,.
\end{equation}
Notice that the sign of the anticommutator is changed when compared to the Hermitian 
case discussed, for instance, in \cite{trisusy1,trisusy2}. 
As the intertwining operators $BA$ and $A^\sharp B^\sharp$ commute with $PT$, this 
superalgebraic structure can be identified easily in the system described by $\tilde{H}$ as well,
\begin{equation}\label{susytildeH}
 	[\tilde{\mathcal{L}}_a,\tilde{H}]=0\,,
	\quad \{\tilde{\mathcal{L}}_a\,,
	\tilde{\mathcal{L}}_b\}=-	\delta_{ab}
	(H-E_0)^4\prod_{m=1}^{2n+1}(H-E_n)\,,\quad a,b=1,2\,,
\end{equation}
where the supercharges are defined as $\tilde{\mathcal{L}}_a=BA\mathcal{L}_aA^\sharp B^\sharp$.
 The grading operator of the superalgera (\ref{susytildeH}) is $PT$.
 
Let us make a brief comment on the finite-gap systems where the intermediate Hamiltonian 
satisfies $\breve{H}=H(x+c)$, see (\ref{fpd}) or (\ref{ld}). In these cases, the Lax-Novikov
 operator associated with $\tilde{H}$ reduces to $(2n+3)$-order differential operator 
 $\tilde{\mathcal{L}}(x)=B(x)\mathcal{L}(x+c)B^{\sharp}(x)$. It satisfies $\tilde{\mathcal{L}}^2=
 \prod_{m=1}^{2n+1}(\tilde{H}-E_m)(\tilde{H}-E_0)^2$.

The scheme given
by the relations 
(\ref{susyH}) and (\ref{susytildeH}) corresponds to 
a bosonized supersymmetry, where no fermionic degrees of freedom 
are present in the system
 \cite{bosonized,bosonized1,bosorigin}. 
 However, by using the confluent Darboux-Crum transformations, an 
 extended, exotic supersymmetry can also be constructed. 
 Let us define a matrix Hamiltonian,
\begin{align}\label{Hextended}
\mathcal{H}=\left(
  \begin{array}{cc}
   H & 0 \\ 
    0 & \tilde{H} \\
  \end{array}
    \right),
\end{align}
which resembles the supersymmetric Hamiltonian of the standard supersymmetry, see 
\cite{CKSsusy}. Instead of having linear supercharges, we can construct a
pair of supercharges 
of the second order
based on the confluent Darboux-Crum transformations, 
by taking
the 
Pauli matrix $\sigma_3$, $\sigma_3^2=1$,  as the grading operator,
 \begin{align}\label{fermi1}
\mathcal{Q}_1=\left(
  \begin{array}{cc}
   0 & A^\sharp B^\sharp \\ 
    B A & 0 \\
  \end{array}
    \right), \qquad \mathcal{Q}_2=i \sigma_3 \mathcal{Q}_1 \, .
\end{align}
The supercharges commute with the Hamiltonian and anticommute with $\sigma_3$, 
resulting in the following $N=2$ non-linear superalgebra,
 \begin{align}\label{QaQb}
	\left[ \mathcal{H},\mathcal{Q}_a\right]=0\,, \quad 
	\{Q_a, \sigma_3\}=0\,, \quad \{Q_a, Q_b\}=
	2\delta_{ab} (\mathcal{H} -E_0)^2, \quad a,b=1,2 \, .
\end{align} 
The finite-gap nature of the Hamiltonians $H$ and $\tilde{H}$ guarantees 
the
existence of 
integrals $\mathcal{L}$ and $\tilde{\mathcal{L}}$. They can be used to define two bosonic 
supercharges for $\mathcal{H}$,

\begin{align}
\mathscr{L}_1=\left(
  \begin{array}{cc}
   (H-E_0)^2  \mathcal{L} & 0 \\ 
    0 & \tilde{\mathcal{L}} \\
  \end{array}
    \right), \quad \mathscr{L}_2=i\sigma_3 \mathscr{L}_1 \,.
\end{align}
In this definition, the polynomial  $(H-E_0)^4$ was 
inserted to keep the same 
differential
 order $2n+5$ 
of the diagonal elements 
(recall that $\mathcal{L}$ is of order $2n+1$ whereas $\tilde{\mathcal{L}}$ is 
of order $2n+5$). The bosonic and non-linear property of the integrals $\mathscr{L}$ are 
summarized in the following superalgebra,
 \begin{align}\label{LaLb}
	\left[ \mathcal{H},\mathscr{L}_a\right]=0\,, 
	\quad [\mathscr{L}_a, \sigma_3]=0\,, 
	\quad \{\mathscr{L}_a, 
	\mathscr{L}_b\}=2\delta_{ab} (\mathcal{H}-E_0)^4\prod_{m=1}^{2n+1}(\mathcal{H}-E_m)
	, \quad a,b=1,2 \, .
\end{align} 
The full supersymmetric structure gets enlarged if we consider also a new set of 
fermionic integrals by multiplying the integrals (\ref{fermi1}) by the extended 
Lax-Novikov operator $\mathscr{L}_1$,
\begin{align}
	\mathcal{S}_1=\mathcal{Q}_1 \mathscr{L}_1\,, 
	\qquad \mathcal{S}_2=i \sigma_3 
	\mathcal{Q}_1 \mathscr{L}_1\,,
\end{align}
which satisfy
\begin{align}\label{HSasig}
	\left[ \mathcal{H},\mathcal{S}_a\right]=0\,, \qquad \{S_a, \sigma_3\}=0\,.
	\end{align}
Finally, the nonlinear superalgebra can be completed by means of the remaining 
(anti)commutation relations,
\begin{align}\label{sa2}
	[ \mathcal{Q}_a, \mathscr{L}_2]&= -2 \epsilon_{ab} \mathcal{S}_{b}\,, 
	\qquad [ 
	\mathcal{S}_a, \mathscr{L}_2]= -2 \epsilon_{ab} \prod_{m=1}^{2n+1}(
	\mathcal{H}-E_m)(\mathcal{H}-E_0)^4 \mathcal{Q}_{b}\,, \\ 
	\{  \mathcal{Q}_a, \mathcal{S}_b \} &=2 \mathcal{S}_a \mathcal{Q}_b+2\delta_{ab}
	 (\mathcal{H} -E_0)^2\mathscr{L}_1\,.\label{sa2+}
\end{align}

Hence, the extended system (\ref{Hextended}) 
is characterized by exotic supersymmetric structure described 
by the four supercharges $\mathcal{Q}_a$ and $\mathcal{S}_a$,
and the two bosonic integrals $\mathscr{L}_a$,
which together generate the nonlinear superalgebra
(\ref{QaQb}), (\ref{LaLb}), (\ref{HSasig}), 
(\ref{sa2}), (\ref{sa2+}).

\section{Discussion}
We showed that the confluent Darboux-Crum
(double-step Darboux-Jordan) 
transformation is an efficient tool for creating $PT$-symmetric 
systems that are asymptotically real and periodic, 
and have a periodicity defect disappearing for large $|x|$.
We pointed out the
difference between the confluent and usual second-order Darboux-Crum 
transformations. 
Whereas the intertwining operators of the standard one annihilate two 
(formal) eigenstates of the 
initial
Hamiltonian, the intertwining operators of the generalized 
transformation annihilate, besides an eigenstate of $H$, 
also the associated Jordan state, see Eqs.
 (\ref{A28}), (\ref{A26}), (\ref{A12}) and (\ref{A15}).

The described confluent Darboux-Crum
 transformations were applied to 
 a
 generic Hamiltonian with real even periodic
  potential (\ref{eq1}). We discussed the general aspects of the construction 
like existence of (quadratically integrable) bound 
states and asymptotic behavior of the created systems. We showed
 that 
  bound states
   in the continuous 
   part of the
     spectrum are associated with invisibility of the periodicity defects.  

It was also showed that the decay rate of the periodicity defect is determined by the position of the
bound-state
  energy in the spectrum; when it belongs to the interior of the energy band or it can be
  identified with the band-edge energy, amplitudes of the defects disappear as $x^{-1}$ or $x^{-2}$.
   When the bound state corresponds to discrete energy in the 
   energy gap or lower forbidden band,  exponential decay 
   of the defect takes place. It would be interesting to verify whether this observation is of general validity, i.e. whether
a periodicity defect of  the $1/x$ or $1/x^{2}$ decay 
induces a bound state in the energy continuum, 
whereas the defects with exponential decay are
responsible for discrete energies.

The application of the generic results was focused on the special class of 
periodic systems that
 possess finite number of gaps in their spectra. In the subsection \ref{FP}, we considered 
reflectionless systems derived by the confluent Darboux-Crum
 transformation from the 
free-particle model.  In particular, we constructed
reflectionless $PT$-symmetric systems which are \textit{completely} invisible.

 In the subsection \ref{LAMEsection}, we discussed asymptotically periodic systems with
  periodicity defects derived from the one-gap system described by the Lam\'e equation. 
  It is worth noticing that in the recent work \cite{longwork}, similar systems 
  with 
  soliton
   defects described by a \textit{Hermitian} 
   Hamiltonian were constructed from the  one-gap
   Lam\'e system using 
   standard 
   Darboux-Crum transformations. 
There, an analysis was provided how to construct new 
Hamiltonians with periodicity defects, that induce bound states 
with discrete energy in the gap or in the lower forbidden band, below the valence band.
 The subsection \ref{LAMEsection} extends those results with the use
  of the generalized (confluent)
  Darboux-Crum transformations. It allowed us
to construct systems with bound state energy of arbitrary value. In particular, we constructed
 explicit models where the bound state was associated with the band-edge energy. We also 
 discussed the specific case where the intermediate Hamiltonian coincides with the original 
 one up to a complex shift of coordinates.

 In section \ref{hifi}, we discussed a set of integrals of 
 motion associated with finite-gap systems,
 which give rise to bosonized and exotic supersymmetries.
 We showed in detail how the superalgebraic structure of the integrals of motion differs 
 from the standard case  when the supercharges based on the confluent 
 Darboux-Crum transformation are taken into account. 
 
Despite illustrating our results on the finite-gap systems, we would like to stress that the results
 of section~\ref{PTGENsection} 
  apply to a broad class of real even  periodic potentials.

The invisibility of periodicity defects discussed in this paper resembles the Klein tunneling. 
This phenomenon occurs in relativistic quantum mechanics where spin-$1/2$ particles can 
tunnel through strong electrostatic barriers. When the particles are massless, the barrier 
becomes invisible for the particles in the sense that it has no effect on their dynamics, 
independently on its actual form. Although Klein tunneling was not observed for elementary 
particles, it is manifested in carbon nanostructures where dynamics is governed by one- or 
two-dimensional Dirac equation. There, it can cause absence of backscattering of Dirac 
fermions on impurities in carbon nanotubes, see e.g. \cite{KleinFP} and references therein.   
In our case, the invisibility is very model-sensitive. On the contrary to the Klein effect, even 
a slight change of the potential can make it visible as the beam or particles start to scatter 
off it. The physics behind these two phenomena is distinct; whereas Klein tunneling stems 
on existence of solutions corresponding to antiparticles, invisibility discussed in this paper
 is rather the result of a fine interference of transmitted and 
 reflected waves on the particular potential barriers.


\newpage

{\bf Acknowldegments} 
FC thanks M. Tudorovskaya for useful discussions and for 
her assistance and explanations about numerical methods.
He is 
supported by the Alexander von Humboldt Foundation. 
VJ was supported by the GA\v CR Grant No.15-07674Y. 
FC 
and MP
thank Nuclear Physics Institute of Czech Republic, 
where a part of this work was done, for hospitality. 
VJ is grateful for warm hospitality of the Universidad 
de Santiago de Chile. He is also grateful for warm 
hospitality that he experienced in CECs. CECs is funded
 by the Chilean Government through the Centers of Excellence Base
Financing Program of CONICYT.
We also acknowledge partial financial support from 
FONDECYT Grants No. 1130017 and 11121651,  CONICYT  Grant No. 79112034 (Chile), 
and Proyecto Basal USA 1298.

\underline{}



\begin{thebibliography}{99}
\bibitem{Milonni} P. W. Milonni, in Coherence Phenomena in Atoms and Molecules in Laser 
Fields, edited by A. D. Bandrauk and S. C. Wallace (Plenum, New York, 1992), p. 45. 

\bibitem{eo1} 
A. Guo, G. J. Salamo, D. Duchesne, 
R. Morandotti, M. Volatier-Ravat, V. Aimez, G. A. Siviloglou,
 and D. N. Christodoulides, 
 Phys.\ Rev.\ Lett.\  {\bf 103}, 093902 (2009).

\bibitem{eo2} 
C. E. R\"uter, K. G. Makris, R. El-Ganainy, D. N. Christodoulides, M. Segev, and D. Kip, 
Nat.\ Phys {\bf 6}, 192 (2010).

\bibitem{eo3} 
A. Regensburger, C. Bersch, M.-A. Miri, G. Onishchukov, D. N. Christodoulides, and U. Peschel, 
Nature {\bf 488}, 167 (2012).

\bibitem{po1} 
R. El-Ganainy, K. G. Makris, D. N. Christodoulides, and Z. H. Musslimani, 
Opt.\ Lett.\  {\bf 32}, 2632 (2007).

\bibitem{po2} 
K. G. Makris, R. El-Ganainy, D. N. Christodoulides,
 and Z. H. Musslimani, 
 Phys.\ Rev.\ Lett.\  {\bf 100}, 103904 (2008).

\bibitem{po3}  
Z.~H.~Musslimani, K.~G.~Makris, R.~El-Ganainy,
 and D.~N.~Christodoulides,
  Phys.\ Rev.\ Lett.\  {\bf 100}, 030402 (2008).

\bibitem{nono1} 
K. G. Makris, R. El-Ganainy, D. N. Christodoulides,
 and Z. H. Musslimani,   Phys.\ Rev.\ A \  {\bf 81}, 063807 (2010). 

\bibitem{nono2} M. C. Zheng, D. N. Christodoulides, R. Fleischmann, and T. Kottos, 
Phys.\ Rev.\ A \  {\bf 82}, 010103 (2010). 

\bibitem{bp2} 
M. V. Berry, J. Phys. A: Math. Gen. {\bf 31},
 3493 (1998).



\bibitem{ui1}  
 H.~F.~Jones,
  J.\ Phys.\ A {\bf 45}, 135306 (2012) [arXiv:1111.2041 [physics.optics]].
  
\bibitem{ui2} 
A. Mostafazadeh, 
Phys. Rev. A {\bf 90}, 023833 (2014) [arXiv:1407.1760 [quant-ph]]; 
Addendum: Phys. Rev. A {\bf 90}, 055803 (2014).
 
\bibitem{ui3} 
Z. Lin, H.Ramezani, T. Eichelkraut, T. Kottos, H. Cao, and 
D. N. Christodoulides, 
Phys.\ Rev.\ Lett.\  {\bf 106}, 213901 (2011) [arXiv:1108.2493 [physics.optics]].


\bibitem{ui4} 
A. Mostafazadeh, Phys. Rev. A {\bf 91}, 063812 (2015)  [arXiv:1504.01756]
 
\bibitem{i1} 
S. Longhi, J. Phys. A: Math. Theor. {\bf 44}, 
485302 (2011) [arXiv:1111.3448 [quant-ph]]. 
 
\bibitem{i2} 
A. Mostafazadeh, Phys. Rev. A {\bf 87}, 
 012103 (2013) [arXiv:1206.0116 [math-ph]]. 



\bibitem{BR}   
C.~M.~Bender,
  Rept.\ Prog.\ Phys.\  {\bf 70}, 947 (2007) 
  [hep-th/0703096 [HEP-TH]].


\bibitem{MR1}  
 A.~Mostafazadeh,
  Int.\ J.\ Geom.\ Meth.\ Mod.\ Phys.\  {\bf 7}, 1191 (2010) 
  [arXiv:0810.5643 [quant-ph]].


\bibitem{MR2} A.~Mostafazadeh,
  Phys.\ Scripta {\bf 82}, 038110 (2010)
  [arXiv:1008.4680 [quant-ph]].
 

\bibitem{benderZnojilmostafazadeh1}
A.~Mostafazadeh, 
 J.\ Phys.\ A {\bf36}, 7081 (2003), [arXiv:quant-ph/0304080];

 \bibitem{benderZnojilmostafazadeh2}
C.~M.~Bender, J.~Brod, A.~Refig, and M.~Reuter,
J.\ Phys.\ A {\bf37}, 10139 (2004) [arXiv:quant-ph/0402026];

\bibitem{benderZnojilmostafazadeh3}
M. Znojil, 
Rendic.\ Circ.\ Mat.\ Palermo, Ser. II, Suppl. {\bf 72}, 211 
(2004)
[arXiv:math-ph/0104012].


\bibitem{bp1}   
S.~Klaiman, U.~Gunther, and N.~Moiseyev,
  Phys.\ Rev.\ Lett.\  {\bf 101}, 080402 (2008), [arXiv:0802.2457 [quant-ph]].

 
 \bibitem{bp3} 
 E.~M.~Graefe and  H. F. Jones,   Phys. Rev. A {\bf 84}, 013818 (2011) 
 [arXiv:1104.2838 [physics.optics]].

\bibitem{Bender:2002vv} 
  C.~M.~Bender, D.~C.~Brody and H.~F.~Jones,
  Phys.\ Rev.\ Lett.\  {\bf 89}, 270401 (2002)
  [Phys.\ Rev.\ Lett.\  {\bf 92}, 119902 (2004)]
  [quant-ph/0208076].


\bibitem{CKSsusy}  
F.~Cooper, A.~Khare, and U.~Sukhatme,
  Phys.\ Rept.\  {\bf 251}, 267 (1995), [arXiv:hep-th/9405029].
 

\bibitem{trisusy1}
F.~Correa, V.~Jakubsk\'y, L.~M.~Nieto, and M.~S.~Plyushchay,
  Phys.\ Rev.\ Lett.\  {\bf 101}, 030403 (2008)
  [arXiv:0801.1671 [hep-th]];

\bibitem{trisusy2}
F.~Correa, V.~Jakubsk\'y, and M.~S.~Plyushchay,
  J.\ Phys.\ A {\bf 41}, 485303 (2008)
  [arXiv:0806.1614 [hep-th]].


\bibitem{exotic1}
  M.~S.~Plyushchay and L.~M.~Nieto,
  Phys.\ Rev.\ D {\bf 82}, 065022 (2010),
  [arXiv:1007.1962 [hep-th]].


\bibitem{exotic2}
M.~S.~Plyushchay, A.~Arancibia, and L.~M.~Nieto,
  Phys.\ Rev.\ D {\bf 83}, 065025 (2011)
  [arXiv:1012.4529 [hep-th]].

 
\bibitem{MPAA2013}   
A.~Arancibia, J.~M.~Guilarte and M.~S.~Plyushchay,
  Phys.\ Rev.\ D {\bf 88}, 085034 (2013)
  [arXiv:1309.1816 [hep-th]].

\bibitem{MPAA2014}   
A.~Arancibia and M.~S.~Plyushchay,
  Phys.\ Rev.\ D {\bf 90}, no. 2, 025008 (2014), [arXiv:1401.6709 [hep-th]].
 

\bibitem{Matveev}
V. B. Matveev and M. A. Salle, Darboux Transformations and Solitons
(Springer, Berlin,
1991).

\bibitem{FMPT}   
F.~Correa and M.~S.~Plyushchay,
  Annals Phys.\  {\bf 327}, 1761 (2012)  [arXiv:1201.2750 [hep-th]].


\bibitem{longwork}
A.~Arancibia, F.~Correa, V.~Jakubsk\'y, J.~M.~Guilarte,
 and M.~S.~Plyushchay,
  Phys.\ Rev.\ D {\bf 90}, no. 12, 125041 (2014)
  [arXiv:1410.3565 [hep-th]].

\bibitem{optsusy1} S. M. Chumakov, K. B. Wolf, 
Phys.\ Lett. A {\bf 193}, 51 (1994). 

\bibitem{optsusy2} 
J. Radovanovi\'c, V. Milanovi\'c, Z. Ikoni\'c, and D. 
Indjin, 
Phys. Rev. B {\bf 59}, 5637 (1999).

\bibitem{optsusy3} 
J. Bai and D. S. Citrin, Optics Express {\bf 14}, 4043 (2006). 


\bibitem{optsusy4} 
M.-A. Miri, M. Heinrich, R. El-Ganainy, and D. N. Christodoulides, 
  Phys.\ Rev.\ Lett.\ {\bf 110}, 233902 (2013) [arXiv:1304.6646 [physics.optics]].

\bibitem{optsusy5} 
M. Heinrich, M.-Ali Miri, S. St\"utzer, R. El-Ganainy,
S. Nolte, A. Szameit, and D. N. Christodoulides, 
Nat. Commun. {\bf 5}, 4698 (2014) [arXiv:1401.5734 [physics.optics]].

\bibitem{optsusy6} 
M.-A. Miri, M. Heinrich, and 
D. N. Christodoulides,
Optica {\bf 1}, 89  (2014) [arXiv:1408.0832 [physics.optics]].

\bibitem{optsusy7} 
S. Longhi, 
Optics Letters {\bf 40}, 463 (2015) [arXiv:1411.7144 [physics.optics]].

\bibitem{susyPTopt1} 
S. Longhi and G. Della Valle, 
Europhys. Lett. {\bf 102}, 40008 (2013) [arXiv:1306.0677 [quant-ph]].
 
\bibitem{susyPTopt2}   
M.~A.~Miri, M.~Heinrich,
 and D.~N.~Christodoulides,
  Phys.\ Rev.\ A {\bf 87}, no. 4, 043819 (2013), [arXiv:1305.1689 [physics.optics]].


\bibitem{susyPTopt3} 
S. Longhi and G. Della Valle, 
Annals. Phys. {\bf 334}, 35 (2013) [arXiv:1306.0667 [quant-ph]]. 

\bibitem{susyPTopt4} 
B. Midya, Phys. Rev. A {\bf 89}, 032116 (2014) [ arXiv:1401.4996 [physics.optics]].

\bibitem{susyPTopt5} 
M. Principe, G. Castaldi, M. Consales, A. Cusano, and 
V. Galdi, 
Sci. Rep. {\bf 5}, 8568 (2015). 


\bibitem{susyPTopt6} 
S. Longhi, J. Opt. {\bf 17}, 045803 (2015) [arXiv:1501.02063 [physics.optics]].

\bibitem{NeuWig} 
J.~von~Neumann and E.~Wigner, Phys.\ Z.\ {\bf 30}, 465 (1929).

\bibitem{BICSukhatme}   
J.~Pappademos, U.~Sukhatme,
 and A.~Pagnamenta,
  Phys.\ Rev.\ A {\bf 48}, 3525 (1993) [arXiv:hep-ph/9305336].


\bibitem{Pursey} 
T. A. Weber and D. L. Pursey, Phys. Rev. A {\bf 50}, 4478 (1994).

 
\bibitem{CSUSY1}
David J Fern\'andez and Encarnaci\'on 
Salinas-Hern\'andez, 
J. Phys. A: Math. Gen. \textbf{36}, 2537 (2003)

\bibitem{CSUSY2}
D.~J.~Fern\'andez C. and E.~Salinas-Hern\'andez,
  Phys.\ Lett.\ A {\bf 338}, 13 (2005)
  [quant-ph/0502147].


\bibitem{CSUSY3}
C.~D.~J.~Fern\'andez and E.~Salinas-Hern\'andez,
  J.\ Phys.\ A {\bf 44}, 365302 (2011)
  [arXiv:1105.2333 [quant-ph]].
 

\bibitem{CSUSY4}
A. Schulze-Halberg, 
Eur. Phys. J. Plus {\bf 128}, 68 (2013)

\bibitem{CSUSY5} 
  A.~Contreras-Astorga and A.~Schulze-Halberg,
  Annals Phys.\  {\bf 354}, 353 (2015).

  \bibitem{cc1} 
  H.~C.~Rosu, S.~C.~Mancas and P.~Chen,
  Annals Phys.\  {\bf 343}, 87 (2014).


\bibitem{cc2} 
  H.~C.~Rosu, S.~C.~Mancas and P.~Chen,
  Annals Phys.\  {\bf 349}, 33 (2014)
  [arXiv:1311.6866 [math-ph]].
  

\bibitem{BICoptics3} 
N. Prodanovi\'c, V Milanovi\'c, and J Radovanovi\'c, 
J. Phys. A: Math. Theor. {\bf 42},
 415304 (2009).


\bibitem{BICoptics4}
 J.S. Petrovi\'c, V. Milanovi\'c, and  Z. Ikoni\'c,
 Phys. Lett. A  {\bf 300}, 595 (2002).

\bibitem{BICoptics1} 
D. C. Marinica, A. G. Borisov, and S. V. Shabanov
Phys. Rev. Lett. {\bf 100}, 183902 (2008).

\bibitem{BICoptics2} 
E. N. Bulgakov and A. F. Sadreev, Phys. Rev. {\bf B} 78, 075105 (2008).

\bibitem{BICoptics5} 
S.~Longhi,
  Opt.\ Lett.\  {\bf 39}, 1697 (2014)
  [arXiv:1402.3761 [quant-ph]].

 \bibitem{susypt1}
  F.~Cannata, G.~Junker and J.~Trost,
  Phys.\ Lett.\ A {\bf 246}, 219 (1998), [arXiv:quant-ph/9805085].

\bibitem{BDM}
  C.~M.~Bender, G.~V.~Dunne and P.~N.~Meisinger,
  Phys.\ Lett.\ A {\bf 252}, 272 (1999),
  [arXiv:cond-mat/9810369].

\bibitem{jones}
H.F Jones, Phys. Lett. A {\bf 262}, 242 (1999).

\bibitem{cervero}
J. M. Cerver\'o,
Phys. Lett. A {\bf 317}, 26 (2003).

\bibitem{shin}
K.C. Shin,
J. Phys. A: Math. Gen. {\bf 37}, 8287 (2004), [arXiv:math-ph/0404015].

\bibitem{samsonovroy}
B. F. Samsonov and  P. Roy,
  J. Phys. A: Math. Gen. {\bf 38}, L249 (2005), [arXiv:quant-ph/0503040].


\bibitem{Megnus} 
W. Magnus and  S. Winkler,
\emph{ Hill's equation} (Wiley, New York, 1966).  


\bibitem{MP2}
S.~M.~Klishevich and M.~S.~Plyushchay,
  Nucl.\ Phys.\ B {\bf 606}, 583 (2001)
  [arXiv:hep-th/0012023].


\bibitem{ZnojilPT}
M. Znojil, 
J.\ Phys.\ A {\bf34}, 9585 (2001) [arXiv:math-ph/0102034].

\bibitem{MFPT}
F.~Correa and M.~S.~Plyushchay,
  Phys.\ Rev.\ D {\bf 86}, 085028 (2012)
  [arXiv:1208.4448 [hep-th]].
 

\bibitem{drazin}
P. Drazin and 
R. Johnson, \emph{Solitons: An Introduction} 
(Cambridge University Press, Cambridge,
England, 1996) (see Q1.12, p.18).


\bibitem{ZnojilCalogero1}   
M.~Znojil and M.~Tater,
  J.\ Phys.\ A {\bf 34}, 1793 (2001) [arXiv:quant-ph/0010087].

   
\bibitem{ZnojilCalogero2}
A.~Fring, M.~Znojil, 
J.\ Phys.\ A {\bf 41}, 194010 (2008) [arXiv:0802.0624 [quant-ph]]. 

\bibitem{mateev}
V. B. Matveev and M. A. Salle, 
\emph{Darboux Transformations and Solitons} (Springer, Berlin,
1991).


 \bibitem{BC1}
J.L. Burchnall and T.W. Chaundy,
Proc. London Math. Soc. Ser. 2, {\bf 21}, 420 (1923).


\bibitem{BC2}
J.L. Burchnall, T.W. Chaundy,
Proc.  Royal Soc. London A {\bf 118}, 557  (1928).


\bibitem{Ince}
E.L. Ince,
{\it Ordinary differential equations},
(Dover, 1956).


\bibitem{Krich}
I.M. Krichever, 
Funct. Anal. Appl. {\bf 12}, 175 (1978).


\bibitem{bosonized}
  M.~S.~Plyushchay,
  Annals Phys.\  {\bf 245}, 339 (1996)
  [arXiv:hep-th/9601116].
  
 \bibitem{bosonized1}
M.~S.~Plyushchay,
  Int.\ J.\ Mod.\ Phys.\ A {\bf 15}, 3679 (2000)
  [arXiv:hep-th/9903130].
  
  \bibitem{bosorigin}
    V.~Jakubsk\'y, L.~M.~Nieto and M.~S.~Plyushchay,
  Phys.\ Lett.\ B {\bf 692} (2010) 51
  [arXiv:1004.5489 [hep-th]].
  
\bibitem{KleinFP}
V.~Jakubsk\'y, L.~M.~Nieto and M.~S.~Plyushchay,
  Phys.\ Rev.\ D {\bf 83}, 047702 (2011)
  [arXiv:1010.0569 [cond-mat.mes-hall]].


\end{thebibliography}
\end{document}